\documentclass[aps,prb,reprint,twocolumns,amsmath,amssymb]{revtex4-1}
\usepackage[]{graphicx}
\usepackage{dcolumn}
\usepackage{bm}
\usepackage{color}

\newcommand{\ibox}[1]{#1}
\begin{document}
\title{Effect of anisotropy on small magnetic clusters}
\author{Alfred Hucht$^1$, Sanjubala Sahoo$^1$, Shreekantha Sil$^2$, and Peter Entel$^1$}
\affiliation{$^1$Faculty of Physics and Center for Nanointegration CeNIDE,
University of Duisburg--Essen, 47048 Duisburg, Germany \\
$^2$Department of Physics, Visva Bharati University, Santiniketan, 
731235 West Bengal, India}
\date{August 31, 2011}
\begin{abstract}
The effect of dipolar interaction and local uniaxial anisotropy on the 
magnetic response of small spin clusters where spins are located
on the vertices of icosahedron ($I_h$), cuboctahedron ($O_h$), 
tetrahedron ($T_h$) and square geometry have been investigated. 
We consider the ferromagnetic and antiferromagnetic spin-$1/2$ and 
spin-1 Heisenberg model with uniaxial anisotropy and dipolar interaction and apply 
numerical exact diagonalization technique in order to study the 
influence of frustration and anisotropy on the ground state properties 
of the spin-clusters. The ground state magnetization, spin-spin correlation 
and several thermodynamic quantities such as entropy and specific heat are 
calculated as a function of temperature and magnetic field. 
\end{abstract}

\pacs{Valid PACS appear here}
\maketitle

\section{Introduction}
Realizing the promising applications in physics, magnetochemistry and 
biomedicine, molecular magnets have recently been the focal point of intense 
subject of research. Although these materials appear as macroscopic objects, i.e.,
crystals or powders, the intramolecular magnetic interactions are utterly 
negligible as compared to the intramolecular interaction. Thus their magnetic 
properties mainly reflect the ensemble properties of small clusters. It appears that 
in majority of these molecules, the localized single particle magnetic moments couple 
antiferromagnetically and the spectrum is rather well described by the Heisenberg model 
with isotropic nearest neighbor interaction, sometimes augmented by anisotropic terms 
and dipolar interactions~\cite{gatteschi-2006, dorfer-2002, schnack-2000, henderson-2008, rschnalle-2009, seeber-2004, schnack-2010}.
Thus, the interest in the Heisenberg model, which is known for a long time, 
is renewed recently by the successful synthesis of new magnetic clusters 
and magnetic molecules. 

{\it Ab initio} studies show that often these magnetic systems are 
frustrated due to the competing magnetic interactions between the individual 
magnetic moments. The effect of finite sizes, quantum fluctuations and frustrations 
can have dramatic consequences on the energy spectra and can even give rise to new 
phases apart from the conventional N$\acute{e}$el-like order~\cite{konstantinidis-2005, shapira-2002, coffey-1992b,
schnalle-2009,schnack-2010, konstantinidis-2001, rousochatzakis-2008, konstantinidis-2007, schulenburg-2002, 
schroder-2005, schnack-2007, schnack-2010th, slageren-2006, ciftja-1999, ciftja-2000}.
A great deal of effort has been devoted to the theoretical  
studies on magnetic clusters using different theoretical techniques 
to solve the Heisenberg model~\cite{konstantinidis-2005, coffey-1992a, schnalle-2009, honecker-2009}. 

Using exact diagonalization of the antiferromagnetic Heisenberg model,
Konstantinidis {\it et al.}~\cite{konstantinidis-2005} have calculated the ground state
magnetization for a dodecahedron and icosahedron symmetry for $s$ = 1/2 and 1 
and have obtained discontinuity in the field-dependent magnetization
and double peaks in the temperature-dependent specific heat arising due to frustrations.
Using perturbation theory, Coffey {\it et al.}~\cite{coffey-1992a} have studied 
the effect of frustration and connectivity on the magnetic properties of 
a 60-site cluster. Schnalle {\it et al.}~\cite{schnalle-2009} have applied
an approximation of diagonalization scheme to a cuboctahedron for $s$ = 1
and 3/2 in order to obtain the energy spectra. 
In addition to the magnetic properties, several studies exist for the thermodynamic properties of 
clusters. For example, Honecker {\it et al.}~\cite{honecker-2009} have calculated several magneto-thermal 
properties such as the magnetic susceptibility, specific heat and magnetic cooling rate for a cuboctahedron
with different spin quantum numbers using the antiferromagnetic Heisenberg model.
Besides the exact diagonalization method, several other techniques such
as the density matrix renormalization group~\cite{white-1993, schollwock-2005},
cluster expansions~\cite{gelfand-1990}, spin-wave expansions~\cite{mattis-1988, zhong-1993, trumper-2000}
and quantum Monte Carlo techniques~\cite{sandvik-1991, sandvik-1999,engelhardt-2006}
can be used to study the magnetic systems. However, some of these techniques
have drawbacks, for example, quantum Monte Carlo technique
has limitations in describing the systems with geometric frustration.
The advantage of exact diagonalization method relative to these 
approximate methods is that one obtains all informations about 
the whole energy spectra such as the degeneracy, the lowest eigenenergies 
and eigenfunctions from which the ground state as well as finite temperature 
properties can be calculated.

In the present work we have applied the exact diagonalization method
to calculate the properties of clusters with spin-1/2 and 1.
We have studied the magnetic and thermodynamic 
properties of small clusters with the focus on showing the effect of dipolar interaction 
and uniaxial anisotropy on the magnetization behavior in the presence 
of magnetic field, the studies of which are still limited in 
literature~\cite{li-2010, hernandez-2005, li-2008}. In addition, the temperature-dependent 
as well as the ground state spin-spin correlation functions are calculated for these 
clusters and are compared with respect to the classical case. 

The paper is organized as follows: In section\;II, we describe the theoretical 
method used for modeling the quantum clusters. Section\;III discusses the results obtained for 13-atom 
clusters with spin-1/2, including the effect of dipolar interaction. 
Then, section\;IV describes the findings for 4-atom clusters 
with spin-1, where the effect of uniaxial anisotropies and temperature-dependent correlation 
functions has been discussed, and section\;V discusses the results for spin-1 icosahedron in the 
presence of local uniaxial anisotropies. In section\;VI the results are summarized.

\section{Theoretical method}
From first-principles calculations it turns out~\cite{herring-1966} that the interaction between electrons 
may be well represented by a model Hamiltonian describing a set of interacting 
spins $\vec s_{i}$. An important class of such interacting 
spin models consists of spins coupled bilinearly on a finite 
lattice. The Hamiltonian of such a system can be expressed by Heisenberg Hamiltonian,
\begin{equation}\label{eqn:eq1}
 \mathcal H_\mathrm{0} = - \sum_{i < j}J_{i j} \, \vec{s}_{i}\cdot\vec{s}_{j},
\end{equation} 
where in general the sum runs over all pairs. $\vec{s}_{i}$ is the spin operator 
on site $i$ having total spin $s$ and $z$ component of the spin $\vec{s_i}$ can take values 
$s_{i}^{z}=-s, -s+1, \ldots , s$; $J_{ij}$ is the exchange interaction. 
This model describes the ferromagnetic (antiferromagnetic) 
Heisenberg model when $J_{ij}>0$ ($J_{ij}<0$). In one dimension and for only nearest-neighbor couplings 
$J_{ij}=J$, the $s$=1/2 Heisenberg model has been solved analytically by means of the 
Bethe \textit{ansatz}~\cite{bonechi-1992}. Unfortunately, the use of 
the Bethe \textit{ansatz} is quite limited, as this method is only applicable to models in one dimension. 
For higher dimensions, one has to look for approximate methods. When the number of spins 
in the system is small enough, one can solve the problem by employing exact diagonalization 
techniques~\cite{lin-1993}. A straightforward way to study the model Hamiltonian, 
defined in Eq.~(\ref{eqn:eq1}), numerically is simply to obtain the matrix elements of $\mathcal H$ in a basis
of $\vert s^{z}_{1},s^{z}_{2},\ldots,s^{z}_{n} \rangle$,  with the $z$-axis taken as quantization direction,
where $n$ is the total number of spins in the system, and then diagonalize the 
Hamiltonian matrix numerically. 
The Hamiltonian matrix can be decomposed into block structure  
with the use of symmetries of the model. Since the isotropic Heisenberg 
model includes only the scalar product between the spins, the Hamiltonian is rotationally 
invariant in spin space, i.e., it commutes with the square of the total spin of the 
system, $S^2$ and the $z$ component of the total spin, $S^z$. 
Even though it is straightforward 
to work in a $S^{z}$ subspace, there is no simple method to construct 
symmetry adopted eigenstates of $\vec S^2$. Construction of symmetry adopted eigenstates of 
$\vec S^2$ requires more involved 
calculations~\cite{gatteschi-1993, almenar-1999, tsukerblat-2006, schnalle-2010,sinitsyn-2007}. 
Additionally, the Hamiltonian is symmetric under permutations of spins that respect the 
connectivity of our small sized cluster, and
the model possesses time reversal symmetry in the absence of external magnetic 
fields. When we take into account the symmetries in the system, the $S^{z}$ basis states can 
be projected onto states that transform under specific irreducible representation of 
the symmetry group. In this way, the Hamiltonian is block diagonalized into smaller matrices 
and the maximum dimension required for numerical diagonalization is considerably reduced compared to 
full Hilbert space size.

In the presence of an external magnetic field, dipolar interaction and anisotropy 
the Heisenberg Hamiltonian (\ref{eqn:eq1}) is modified to
\begin{equation}\label{eqn:eq2}
\mathcal H = \mathcal H_{0} - B^z S^z + \mathcal H_\mathrm{dipole} + \mathcal H_\mathrm{ani}
\end{equation}
where $B^z$ is the homogeneous external magnetic field defining, without loss of generality, 
the $z$-direction. Here it may be noted that the factor $g\mu_B$ is absorbed into $B^z$  
and $z$ component of the the total spin, $S^{z}=\sum_{i}s_{i}^{z}$, can take values from $-S$ to $S$ in unit steps, 
where $S$ is the maximum total spin of the system. 
The dipolar term $\mathcal H_\mathrm{dipole}$ in Eq.~(\ref{eqn:eq2}) is defined as
\begin{equation}\label{eqn:eq3}
\mathcal H_{\mathrm{dipole}} =
\frac{\mu_0}{4\pi} (g \mu_{\mathrm B})^2
\sum_{i<j}\frac{\vec{s}_i \cdot \vec{s}_j
- 3 (\vec{s}_i \cdot \hat{\vec r}_{ij}) (\hat{\vec r}_{ij} \cdot \vec{s}_j)}{|\vec{r}_{ij}|^3}
\end{equation}
where $\hat{\vec r}_{ij} = \vec{r}_{ij} / |\vec{r}_{ij}|$ is
the unit vector along the line connecting the two spins or dipoles located on the 
sites $i$ and $j$, and the sum runs over all pairs.
$\mathcal H_\mathrm{ani}$ in Eq.~(\ref{eqn:eq2}) represents the local uniaxial anisotropy, which is defined by
\begin{equation}\label{eqn:eq4}
\mathcal H_{\mathrm{ani}} = - \sum_{i} D_i \, (\vec{e_i} \cdot \vec{s_{i}})^{2}
\end{equation}
where $D_i$ are the local uniaxial anisotropy constants and 
$\vec{e_i}$ is the unit vector giving the radial direction from the central spin. 
Since the commutators, $[\mathcal H_\mathrm{dipole},S^z] \ne 0$ and $[\mathcal H_\mathrm{ani},S^z] \ne 0$, 
$S^z$ is no-more a good quantum number in presence of dipolar interaction $\mathcal H_\mathrm{dipole}$ or 
anisotropy term $\mathcal H_\mathrm{ani}$ and 
therefore in presence of dipolar or anisotropy term the block diagonalization 
with respect to different $S^z$ values is not possible.  
 
Now we construct the Hamiltonian matrix in terms of eigenstates of the 
total $S^{z}$ operator and express the Hamiltonian 
in terms of  the raising and lowering operators $s_{i}^{\pm}=s_{i}^{x} \pm i s_{i}^{y}$. 
When $s_{i}^{+}$ and $s_{i}^{-}$ operates on the eigenstates of $s_{i}^{z}$, we have
\begin{equation}
s_{i}^{\pm}\vert s_{i}^{z}\rangle = \sqrt{s(s+1) - s_{i}^{z}(s_{i}^{z} \pm 1)}\,\vert s_{i}^{z}{\pm}1\rangle.
\end{equation}
For example, for a spin-1/2 particle, 
\begin{align*}
s_{i}^{+}\vert \!\!\uparrow_{i}\rangle& = 0,~ s_{i}^{+}\vert \!\!\downarrow_{i}\rangle = \vert \!\!\uparrow_{i}\rangle,\\
s_{i}^{-}\vert \!\!\uparrow_{i}\rangle& = \vert \!\!\downarrow_{i}\rangle,~ s_{i}^{-}\vert \!\!\downarrow_{i}\rangle = 0,
\end{align*}
and for a spin-1 particle,
\begin{align*}
s_{i}^{+}\vert \!\!\uparrow_{i}\rangle& = 0,~ s_{i}^{+}\vert 0_{i}\rangle = \sqrt{2}\vert \!\!\uparrow_{i}\rangle,\\
s_{i}^{+}\vert \!\!\downarrow_{i}\rangle& = \sqrt{2}\vert 0_{i}\rangle,~ s_{i}^{-}\vert \!\!\uparrow_{i}\rangle = \sqrt{2}\vert 0_{i}\rangle,\\
s_{i}^{-}\vert 0_{i}\rangle& = \sqrt{2}\vert \!\!\downarrow_{i}\rangle, s_{i}^{-}\vert \!\!\downarrow_{i}\rangle = 0,
\end{align*}
where in the latter case $\{ \downarrow, 0, \uparrow \}$ denote the three possible values of $s^z_i = -1,0,1$.
In the absence of $\mathcal H_\mathrm{dipole}$ and $\mathcal H_\mathrm{ani}$, the $z$ component of the total spin 
is a conserved quantity and we can decompose 
the Hamiltonian matrix into smaller blocks characterizing each values of the total spin. 
For example, in case of a 13-atom cluster with $s$ = 1/2, the ${2^{13}}\times{2^{13}}$
dimensional Hamiltonian matrix is divided into blocks with 
dimension $\binom{13}{k} \times \binom{13}{k}$, with $k = 0,\ldots,13$. 14 such 
block matrices have to be diagonalized and the largest block matrix 
has $\binom{13}{6} = 1716$ rows. It may be noted that $S^z$ being the good quantum 
number, the Zeeman term is not needed to be included in numerical diagonalization 
process and can be included later by shifting the eigenvalues by $B^zS^z$. However,
these simplification is not possible when dipolar interaction or uniaxial anisotropy 
term is present in the Hamiltonian. 

\begin{figure}
{}\hfill
\ibox{\includegraphics[scale=0.21]{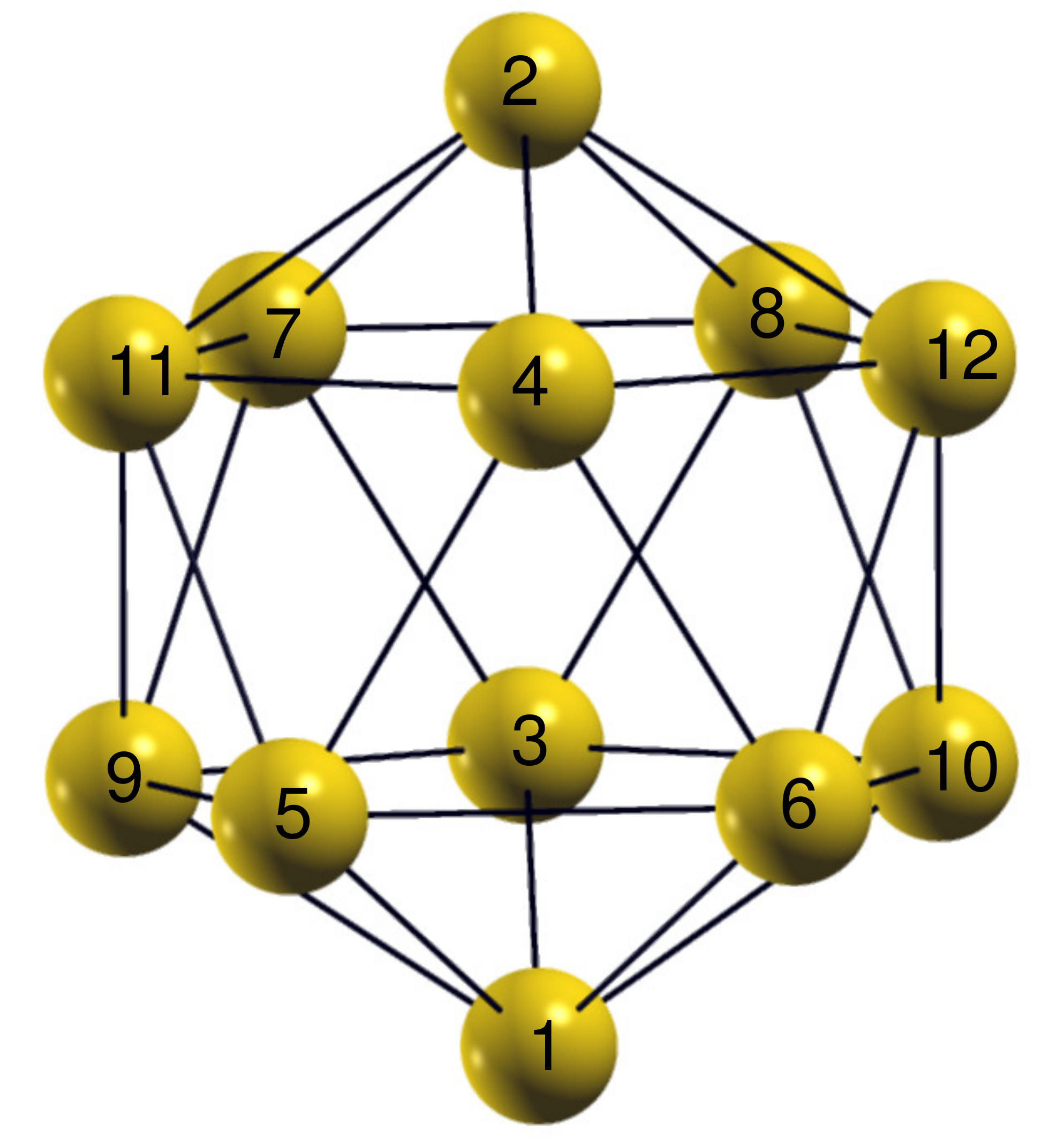}}
\hfill\hfill
\ibox{\includegraphics[scale=0.21]{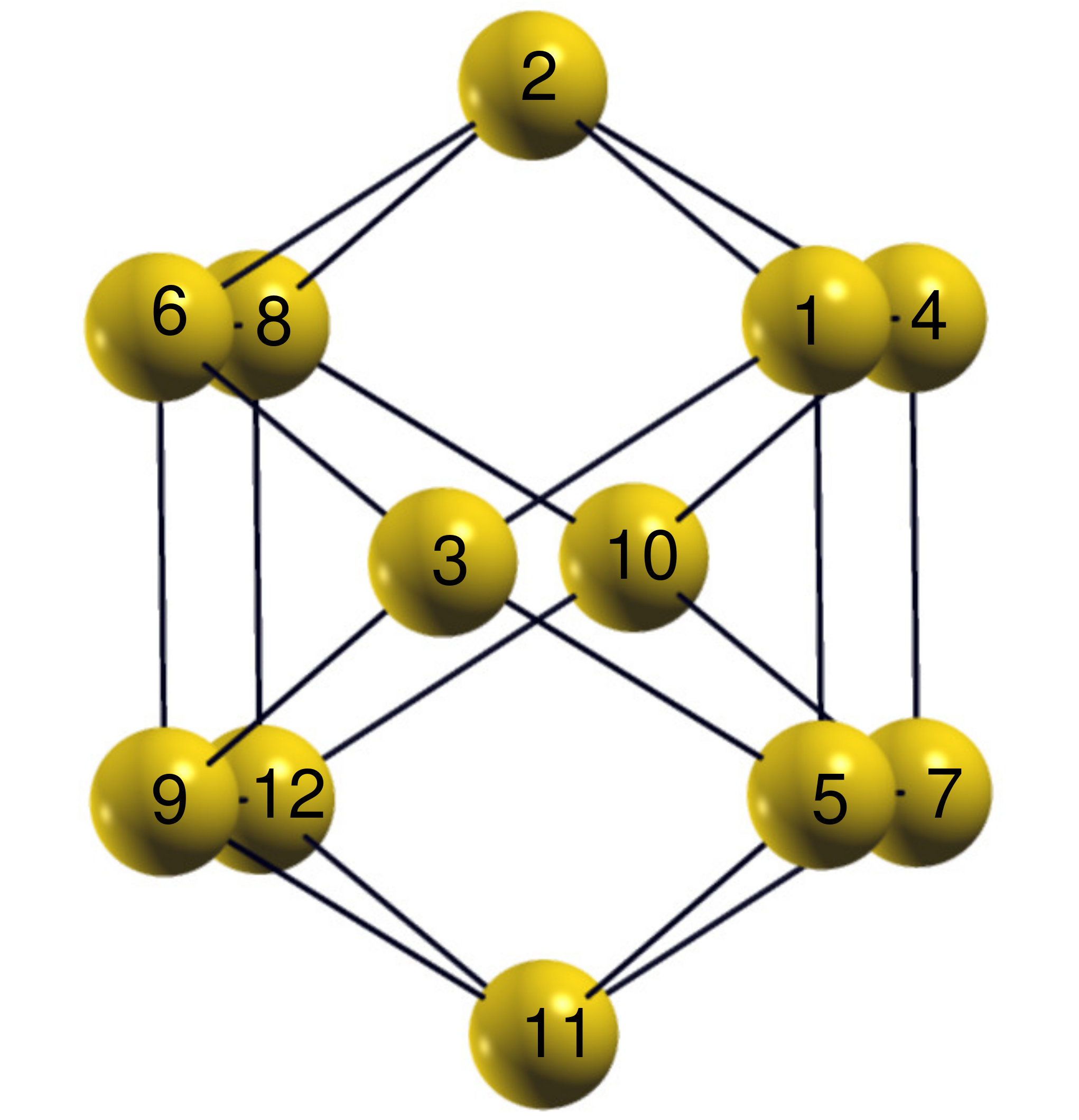}}
\hfill{}
\caption{
\label{fig:ico}(Color online) Schematic picture of ICO (left) and 
CUBO (right) with labeling of each atomic site. Both structures have 12 vertices 
with one atom at center (not shown). The magnetic field is aligned  
parallel to $\vec{r}_2$ for both ICO and CUBO.}
\end{figure}

\section{13-atom clusters with $\boldmath s$ = 1/2}
We have considered two different geometries, namely icosahedron (ICO) and cuboctahedron (CUBO) 
for the investigation of 13-atom clusters. The characteristic feature of the icosahedron, which has 
a connectivity like fullerenes~\cite{kroto-1985, fowler-1995}, is that it possesses 12 vertices, 
20 triangular faces and 30 edges. It is categorized in the symmetry group of $I_{\it h}$, which 
is the point symmetry group with 120 operations~\cite{altmann-1994}. On the other hand, a 
cuboctahedron has 12 vertices with 8 triangular and 6 square faces and 24 identical edges, 
and belongs to the symmetry group $O_{h}$. Schematic pictures of the ICO and 
CUBO geometry are shown in the left and right panel of Fig.~\ref{fig:ico}, respectively. 
Both geometries possess 12 vertices on the surface shell and one atom at the center, 
and can be transformed into
each other via a Mackay transformation~\cite{rollmann-2007}.
Though the number of nearest neighbors for the center atom are same (12) for both geometries, 
each of the surface atoms for both cases possesses different number of nearest 
neighbors, i.e., the ICO has 5 and the CUBO has 4 nearest neighbors in the outer shell. In this
section we first study the ground state properties and thermodynamic quantities 
like entropy and specific heat in the absence of dipolar and uniaxial 
anisotropy term and then switch on the dipolar interaction to investigate its influence 
on ground state properties. 

In the absence of $\mathcal H_\mathrm{dipole}$ and $\mathcal H_\mathrm{ani}$, the total Hamiltonian 
for a 13-atom cluster with nearest-neighbor interaction can be written as
\begin{equation}\label{eqn:eq7}
\mathcal H_{13} =  -J \sum_{\substack{{i,j>0}\\{\langle ij \rangle}}} \vec{s}_{i}\cdot\vec{s}_{j}
-J^{\prime}\sum_{i>0}\vec{s}_{0}\cdot\vec{s}_{i} - B^z \sum_{i}s_{i}^{z}
\end{equation}
where $\vec{s}_0$ is the spin of the center atom, and the first sum runs over all
nearest-neighbor pairs $\langle ij \rangle$ in the surface shell. 
$J$ is the exchange coupling between atoms in the surface shell 
and $J^{\prime}$ is the exchange coupling between central and surface spins.

The energy spectrum for the two clusters namely the ICO and the CUBO clusters can be 
obtained by diagonalizing the above Hamiltonian. In the presence of magnetic field the ground state 
energy is obtained by considering the minimum of the energy eigenvalues from each magnetization sector. 
\begin{figure}
\ibox{\includegraphics[width=\columnwidth]{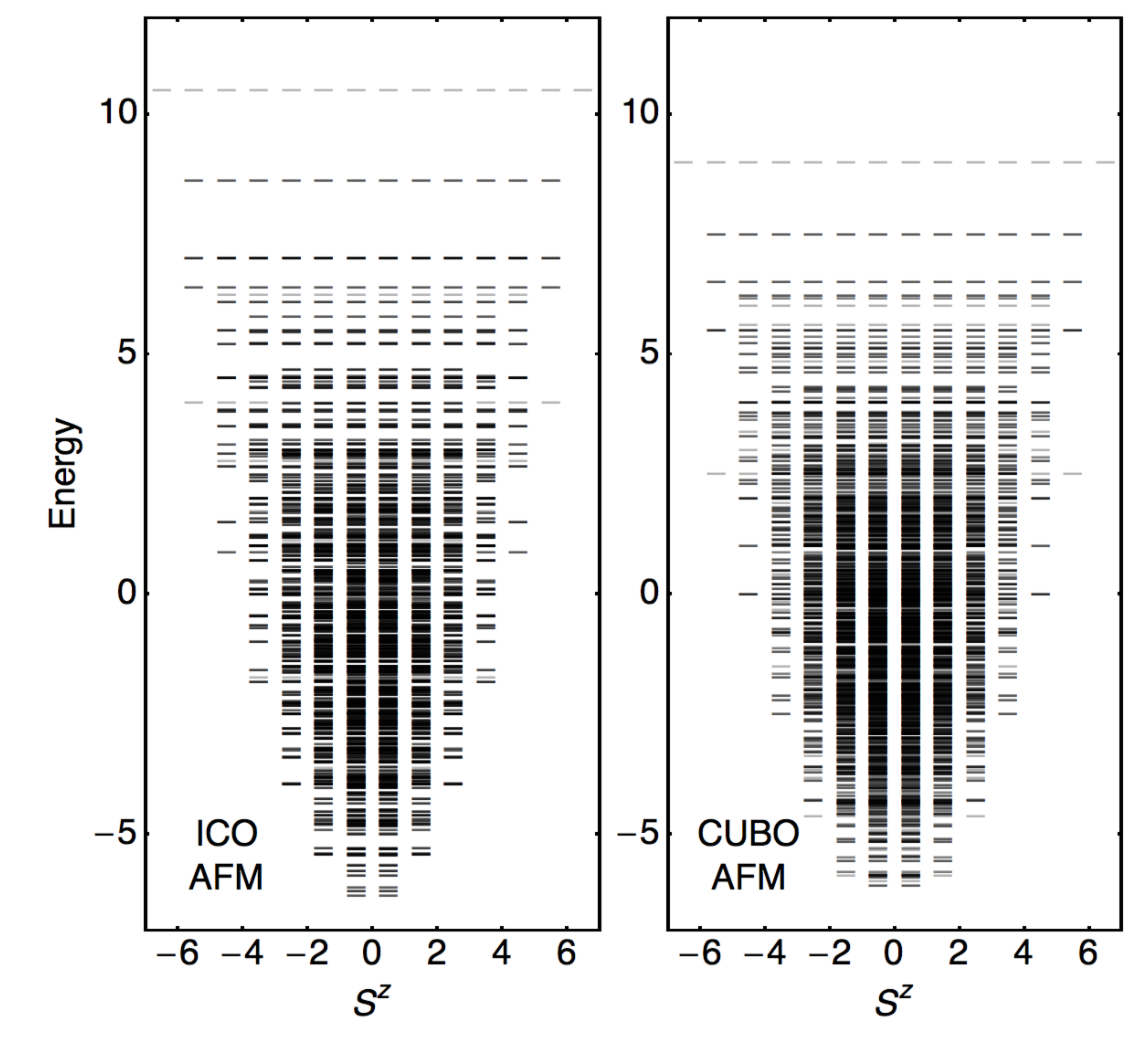}}
\caption{
\label{fig:ff2}All 8192 energy eigenvalues in units of $|J|$ for AFM interactions 
of the 13-atom ICO (left panel) and CUBO (right panel) with spin-1/2. 
The energy levels are shaded according to their degeneracy.
There exists a $\pm{S^z}$ degeneracy in the AFM case for both symmetries. 
For the FM case, the energy spectra are reversed with respect to the 
AFM spectra which fulfills $\mathrm{E_{FM} = -E_{AFM}}$.}
\end{figure}
%
\begin{figure*}
{}\hfill
\ibox{\includegraphics[scale=0.50]{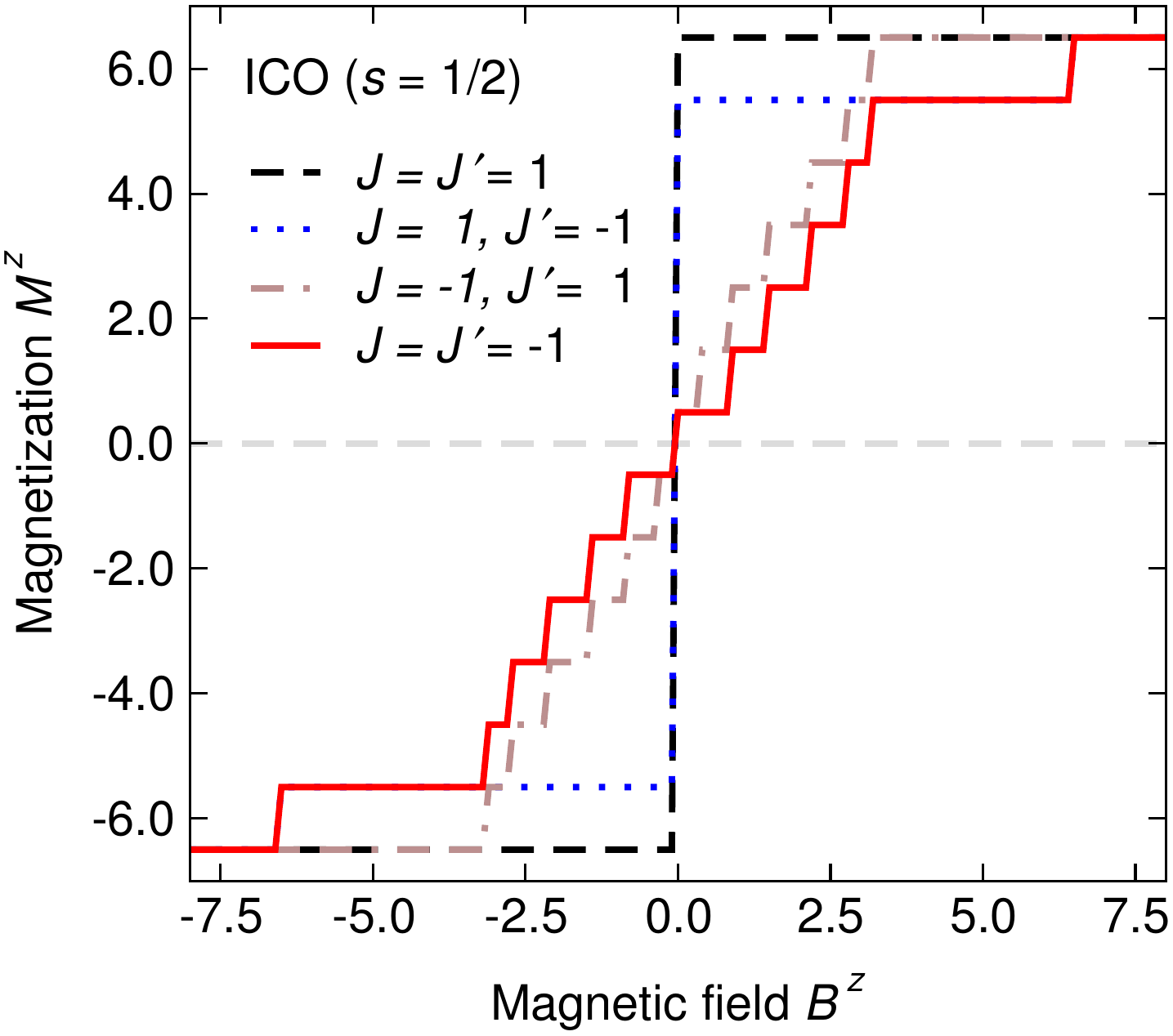}}
\hfill\hfill
\ibox{\includegraphics[scale=0.50]{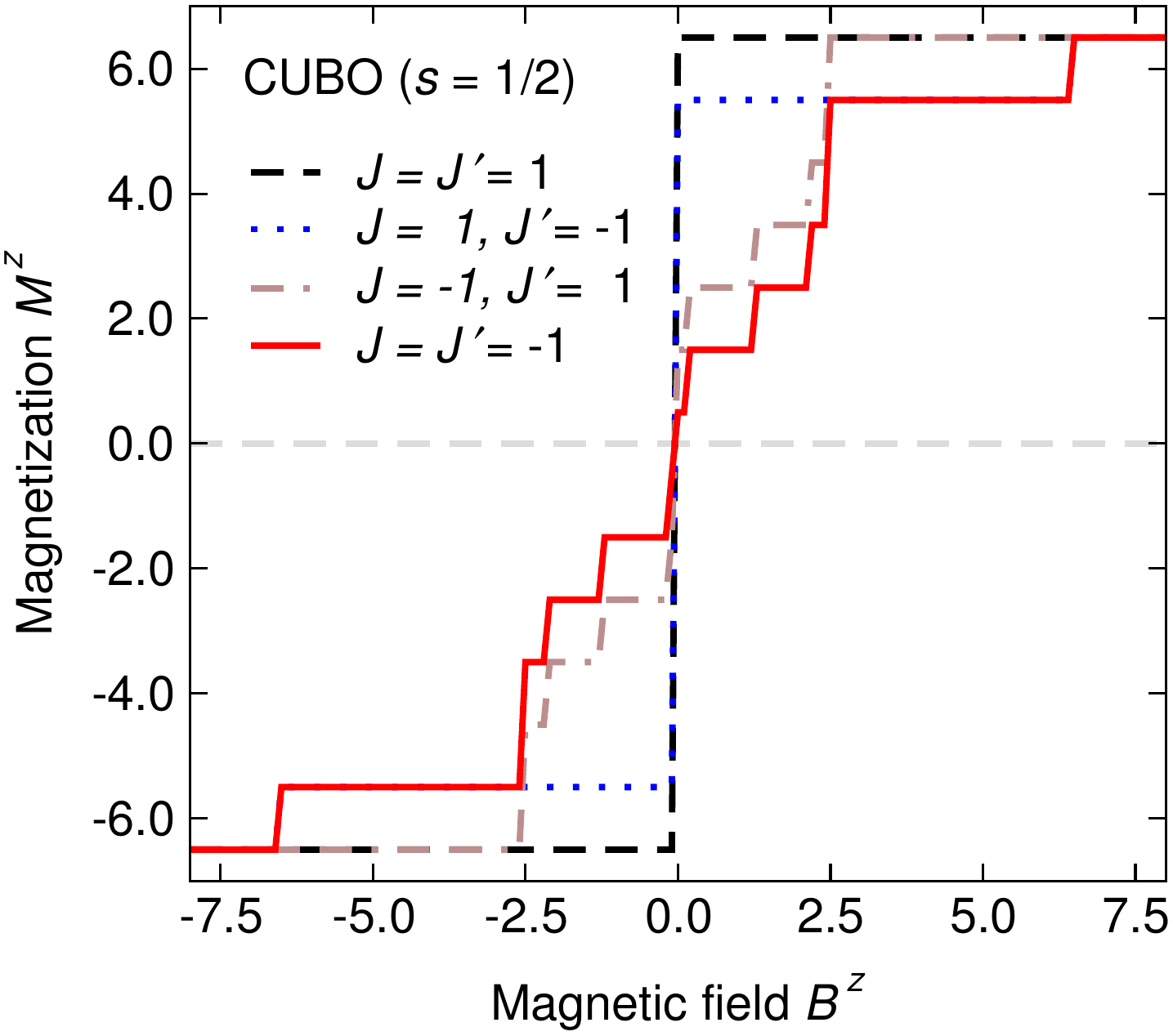}}
\hfill{}
\caption{
\label{fig:ff3} (Color online) Variation of magnetization $M^z$ = $\langle{S^z}\rangle$ 
as a function of the magnetic field for four different exchange interactions as listed in the 
panels of ICO (left panel) and CUBO (right panel). 
$J$ and $J^{\prime}$ are the exchange couplings among the surface spins and center-surface 
spins, respectively. The external magnetic field is measured in units of $|J|$.
}
\end{figure*}
We have calculated the energy eigenvalues for the different exchange couplings defined by (i) all 
spins ferromagnetic ($J = J^{\prime} = 1$), (ii) all spins antiferromagnetic
($J = J^{\prime} = -1$), (iii) central spin is reversed with respect to the ferromagnetic surface ones 
($J = 1$ and $J^{\prime} = -1$), (iv) antiferromagnetic surface spins with ferromagnetic central spin 
($J= -1$ and $J^{\prime} = 1$). However, we will mainly discuss the ferromagnetic ($J = J^{\prime} = 1$) and 
antiferromagnetic ($J = J^{\prime} = -1$) cases. Note that all energies are measured in units of $|J|$ with
$|J|$ is fixed to the value 1 in this work.
\begin{table}[b]
\caption{
\label{tab:table2}Lowest energy eigenvalue $E_0$, degeneracy $K_0$ and lowest energy excitation 
$\Delta E_1 = E_1 - E_0$ for different $S^{z}$ for 13-atom AFM ICO and CUBO. Energies are in units of $|J|$.
}
{
\begin{tabular*}{\columnwidth}{@{\extracolsep{\fill}} ccccccc}
\hline
\hline
\vspace{-0.3cm}\\
$|S^{z}|$ &
$E_0^\mathrm{ICO}$ & $K_0^\mathrm{ICO}$ & $\Delta E_1^\mathrm{ICO}$&
$E_0^\mathrm{CUBO}$ & $K_0^\mathrm{CUBO}$ & $\Delta E_1^\mathrm{CUBO}$\\
\vspace{-0.3cm}\\
\hline
13/2 & $21/2$ & 1 & $-$ & $9$ & 1 & $-$ \\
11/2 & $4$ & 1 & $2.382$ & $5/2$ & 1 & $3$ \\
9/2 & $2-\sqrt 5/2$ & 3 & $0.618$ & $0$ & 5 & $1$ \\
7/2 & $-1.834$\footnote{\scriptsize Zero of $x^3-5x-3=0$} & 5 & $0.102$ & $-5/2$ & 3 & $0.293$ \\
5/2
 & $-3.967$\footnote{\scriptsize Zero of $64 x^6+448 x^5+656 x^4-1184 x^3-3412 x^2-2036 x-53=0$} & 4 & $0.0045$ 
 & $-4.631$\footnote{\scriptsize Zero of $2 x^5+16 x^4+29 x^3-23 x^2-61 x-8=0$} & 1 & $0.339$ \\
3/2
 & $-5.420$\footnote{\scriptsize Zero of 
 $4 x^{10}+84 x^9+700 x^8+2842 x^7+4992 x^6-1726 x^5-21401 x^4-31503 x^3-14082 x^2+4014 x+3402=0$} & 5 & $0.022$ 
 & $-5.869$\footnote{\scriptsize Zero of $64 x^6+960 x^5+4784 x^4+7168 x^3-8148 x^2-19868 x+6361=0$} & 1 & $0.093$ \\
1/2
 & $-6.288$\footnote{\scriptsize Zero of 
 $65536 x^{16} + 2752512 x^{15} + 51707904 x^{14} + 571146240 x^{13} + 4089167872 x^{12}
  + 19595452416 x^{11} + 61510348800 x^{10} + 109531144192 x^9 + 14047096320 x^8 - 488888621568 x^7
  - 1389656886528 x^6 - 2016792866048 x^5 - 1655926247744 x^4 - 669806791648 x^3 - 39673588208 x^2
  + 46200676992 x + 7484904361=0$} & 3 & $0.100$ 
 & $-6.062$\footnote{\scriptsize Zero of 
 $64 x^{21}+3136 x^{20}+70272 x^{19}+952256 x^{18}+8684000 x^{17}+55985680 x^{16}+259611872 x^{15}
 +853909520 x^{14}+1844888624 x^{13}+1761797108 x^{12}-3621087792 x^{11}-18691236512 x^{10}
 -39464764094 x^9-49351650308 x^8-34081746286 x^7-3226424608 x^6+17175800242 x^5+16425687591 x^4
 +6952269434 x^3+1297049762 x^2+47065144 x-4927905=0$} & 3 & $0.093$ \\
\hline
\hline
\end{tabular*}
}
\end{table}

In Table~\ref{tab:table2} we present closed form expressions for the ground state energies $E_0$,
degeneracies $K_0$, as well as the lowest energy gap $\Delta E_1$ for the different $S^z$ sectors 
of ICO and CUBO. The exact polynomials are determined with the \textit{Mathematica}
routine "RootApproximant"~\cite{MMA} using high precision arithmetics with up to 400 digits. 
Due to the two-fold degeneracy for the $\pm{S^z}$ sector, 
where the minimum energy for each positive $S^z$ sector has the same value as that 
of the corresponding negative $S^z$ sector, we have listed the 
results as function of  $|S^z|$ only.    

The whole eigenvalue spectrum in the absence of external magnetic field is depicted 
in Fig.~\ref{fig:ff2} for the AFM interactions of ICO (left panel) and 
CUBO (right panel), where the minimum energy eigenvalues for different magnetization 
of the system are found to be different. More over the energy gaps between the minimum energy eigenvalues of 
the consecutive $S^z$ sectors differ for both ICO and CUBO. This observation identifies the influence 
of symmetry on the nature of eigenvalue spectrum of the system and also explains the nature of the variation of 
the magnetization with respect to the external magnetic field for the two clusters. 
For the FM interaction, we obtain degenerate minimum energies. This occurs 
because of the fact that the Hamiltonian has spin rotational invariance and as a result,  
turning the total spin in another direction does not change the energy of the system. 
\begin{figure*}
\ibox{\includegraphics[scale=0.55]{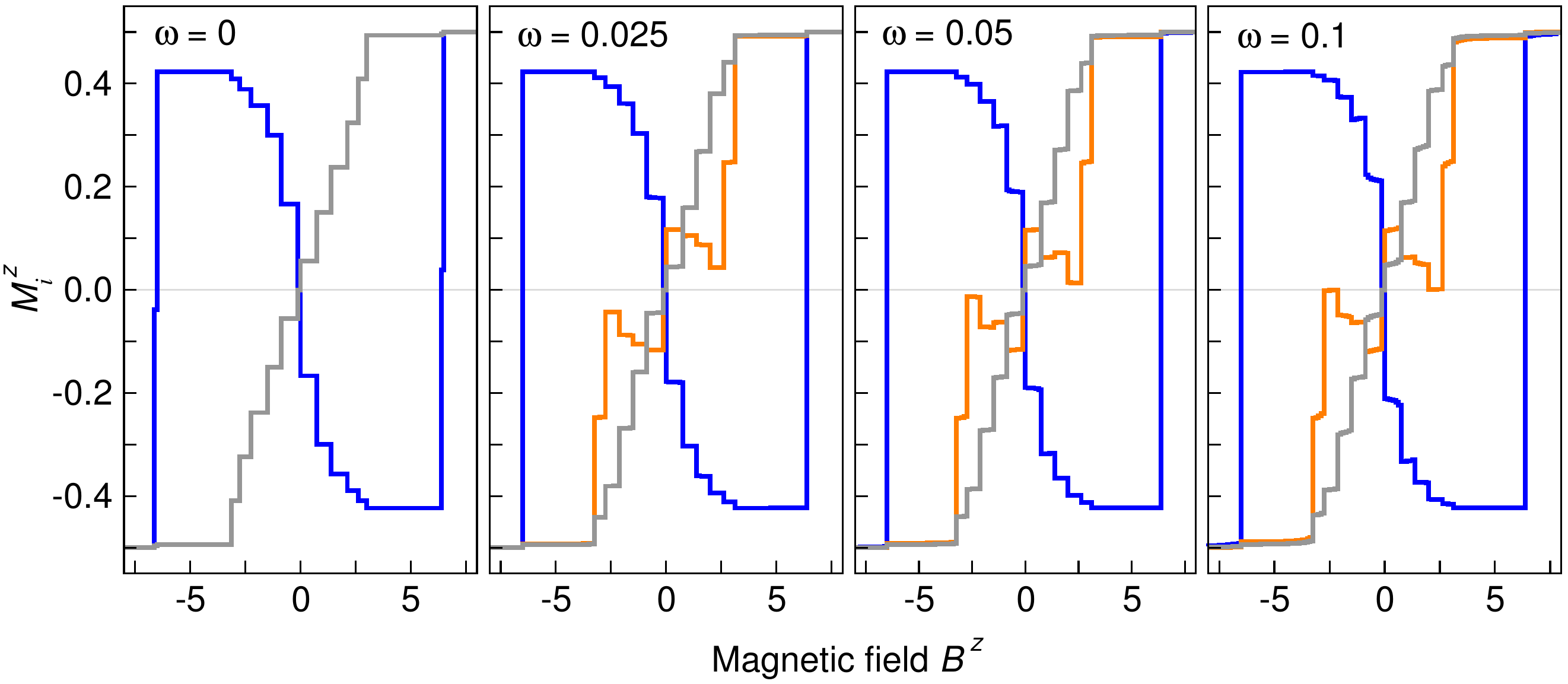}}
\caption{
\label{fig:ff5} (Color online)
Variation of ground state magnetization $M_i^{\mathrm{z}}$ = $\langle s_i^z \rangle$ as a function of 
magnetic field (in units of $|J|$) 
for the AFM case of 13-atom ICO at several values of the reduced dipole coupling strengths $\omega$. 
The dark blue lines in all plots show the field-dependence of magnetization for the 
center spin $\langle s_0^z \rangle$. The orange and light gray lines show the same quantity for the top/bottom atoms 
and remaining 10 atoms on the surface, respectively. The top/bottom spins are strongly affected by dipolar interactions.
}
\end{figure*}

The left panel of Fig.~\ref{fig:ff3} shows the variation of magnetization in the unit of $g \mu_B$ as a function of external 
magnetic field for the four cases of interactions of the ICO (mentioned above). In the presence of an 
external magnetic field, the minimum energy configuration for the AFM 
interaction ($J = J^{\prime} = -1$) gives rise to plateaus, which have been marked by the 
solid red curve in the left panel of Fig.~\ref{fig:ff3}. The appearance of different sizes of 
plateaus is related to the inequivalent energy gaps between the minimum energy values of consecutive $S^z$ sectors. 
On the other hand for FM interaction the ground state energy lies in the 
$S^z=13/2$ sector for all positive values of $B^z$ and in the $S^z = -13/2$ sector 
for all negative values of $B^z$. Thus for the ferromagnetic interaction ($J=J'$), irrespective of the 
values of magnetic field, $|M^z|=13/2$ (see the black dashed line in Fig.~\ref{fig:ff3}).
The right panel of Fig.~\ref{fig:ff3} shows the variation of magnetization 
as a function of magnetic field for the four different set of exchange couplings in the case of  
CUBO geometry. As observed in the case of ICO, a similar behavior for the 
variation of magnetization with respect to the external magnetic field is observed for the FM interaction. 
However, for the AFM interactions of CUBO, the plateaus appearing in the magnetization
have different sizes compared to the ICO, which can be noted from the solid red curve in the right panel
of Fig.~\ref{fig:ff3}. The differences in results is the consequence of the differences in structural symmetries
of the two clusters.
\begin{table}[b]
\caption{
\label{tab:table-dipole}Ground state expectation values of center and surface spins at $\omega = 0$ 
for the AFM case of ICO and the corresponding $S^z$. Note that for $S^z=13/2$ the central spin is oriented 
parallel and Eq.~(\ref{eq:s0si}) does not hold.
}
{
\begin{tabular*}{0.4 \textwidth}{@{\extracolsep{\fill}} ccc}
\hline
\hline
\vspace{-0.3cm}\\
$S^z$ & $\langle s_0^z \rangle$ & $\langle s_i^z \rangle$\\
\vspace{-0.3cm}\\
\hline
 $13/2$ & $+0.5000$  & $0.5000$ \\
 $11/2$ & $-0.4231$ & $0.4936$ \\
 $9/2$ & $-0.4091$  & $0.4091$ \\
 $7/2$ & $-0.3889$  & $0.3241$ \\
 $5/2$ & $-0.3571$  & $0.2381$ \\
 $3/2$ & $-0.3000$  & $0.1500$\\
 $1/2$ & $-0.1667$  & $0.0556$ \\
\hline
\hline
\end{tabular*}
}
\end{table}

\begin{figure*}
\ibox{\includegraphics[scale=0.55]{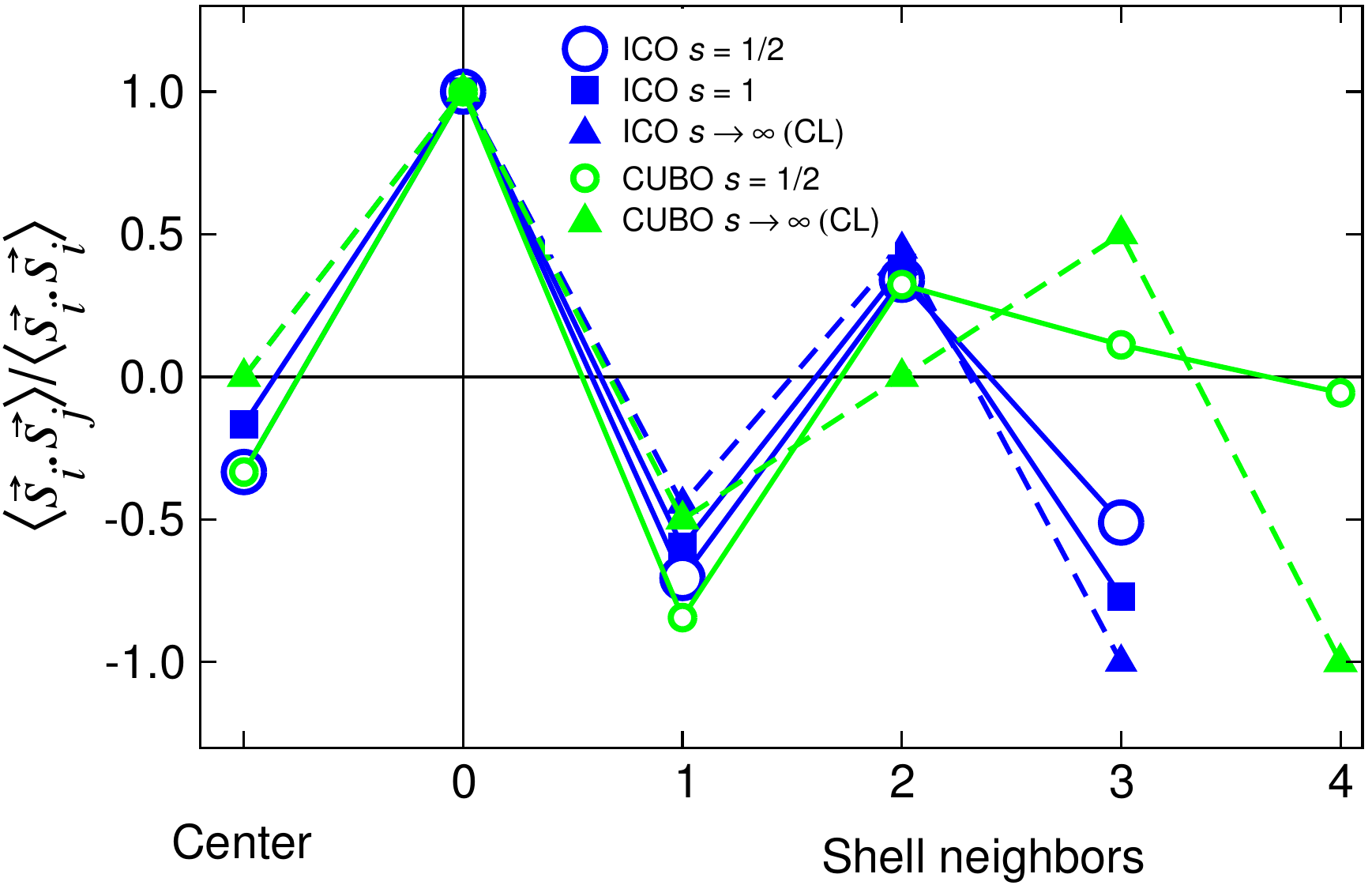}}
\caption{
\label{fig:ffcorr} (Color online)
Ground state correlation functions for $n^{th}$ (with 1 $\le$ $n$ $\le$ 4)
shell neighbors of ICO and CUBO for the AFM case with spin-1/2 
and spin-1. The center and the zero index in the abscissa indicates the correlation functions 
for the center atom and from center to the atom on surface shell, respectively. 
The dashed lines indicate the classical limit for spin-$\infty$ ICO and CUBO.
For the FM case, the correlation functions for ICO and CUBO 
with spin-1/2 possess same magnitude (0.25) for all neighbors. 
}  
\end{figure*}
\begin{figure*}
\ibox{\includegraphics[scale=0.57]{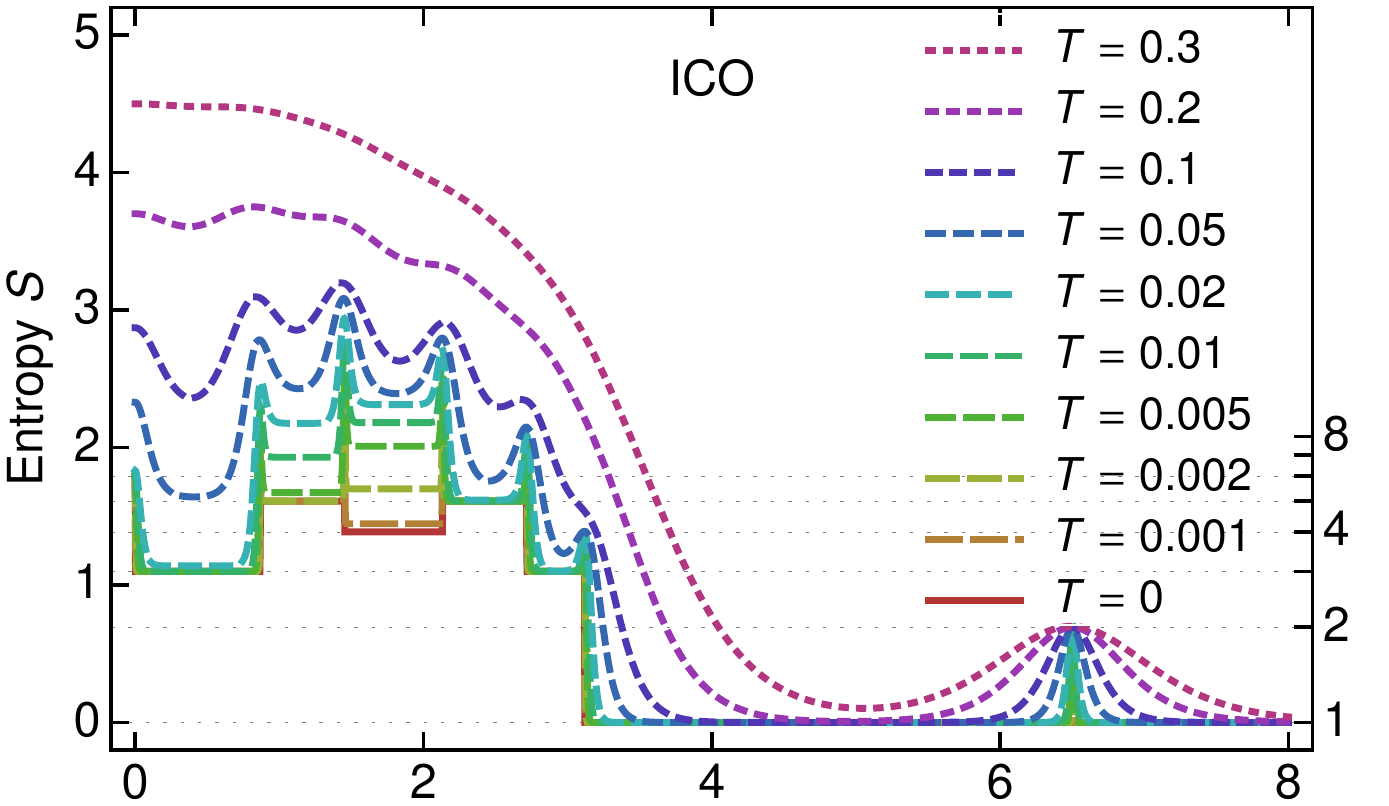}}
\ibox{\includegraphics[scale=0.57]{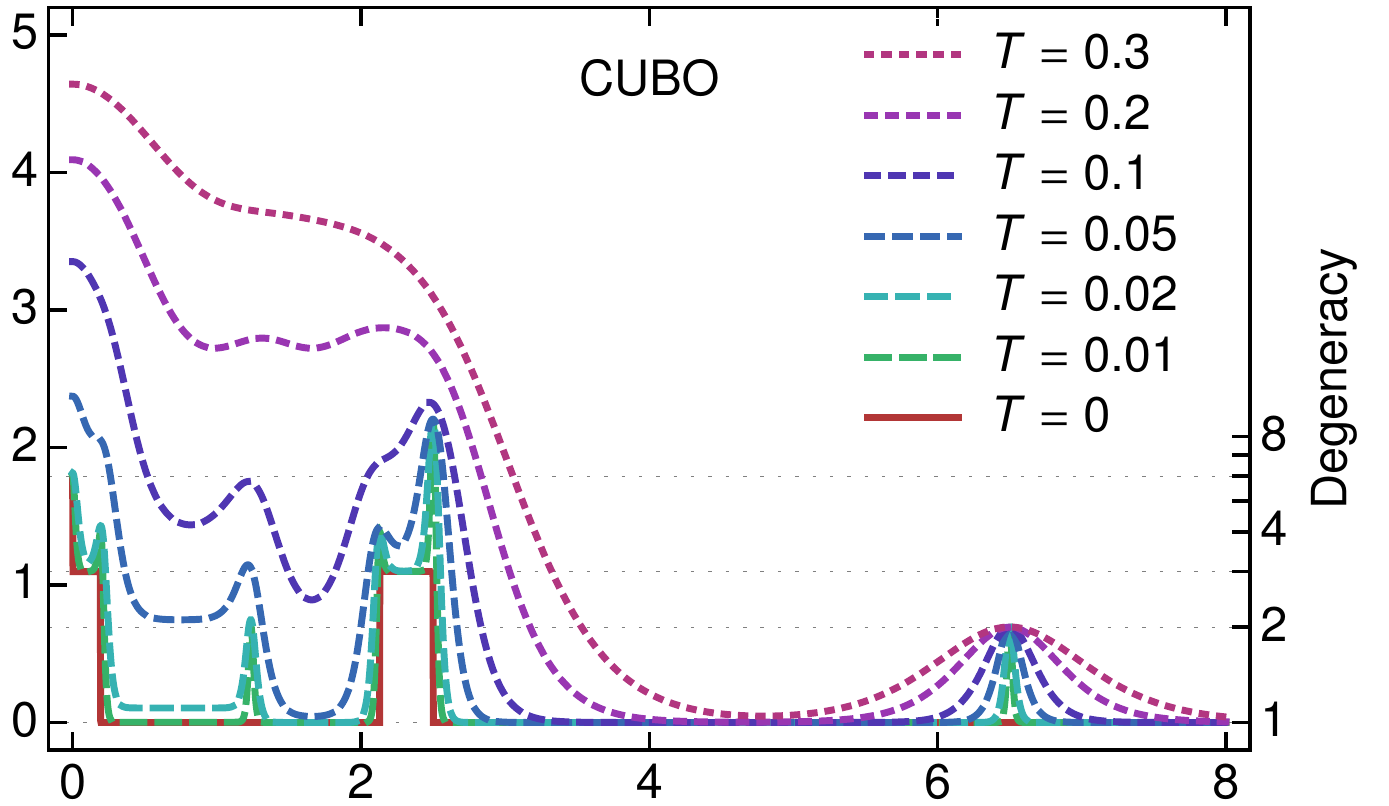}}
\ibox{\includegraphics[scale=0.57]{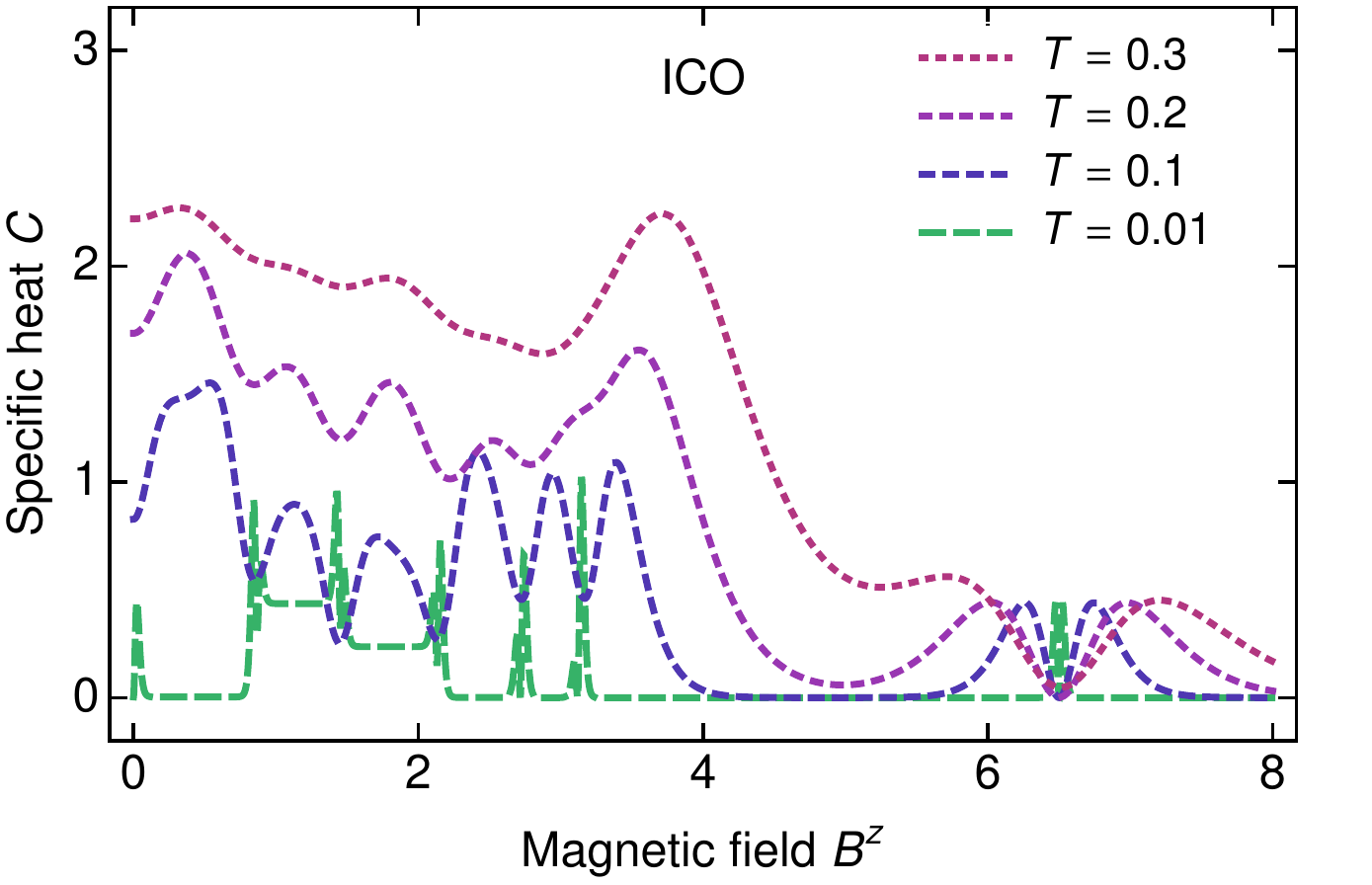}}
\ibox{\includegraphics[scale=0.57]{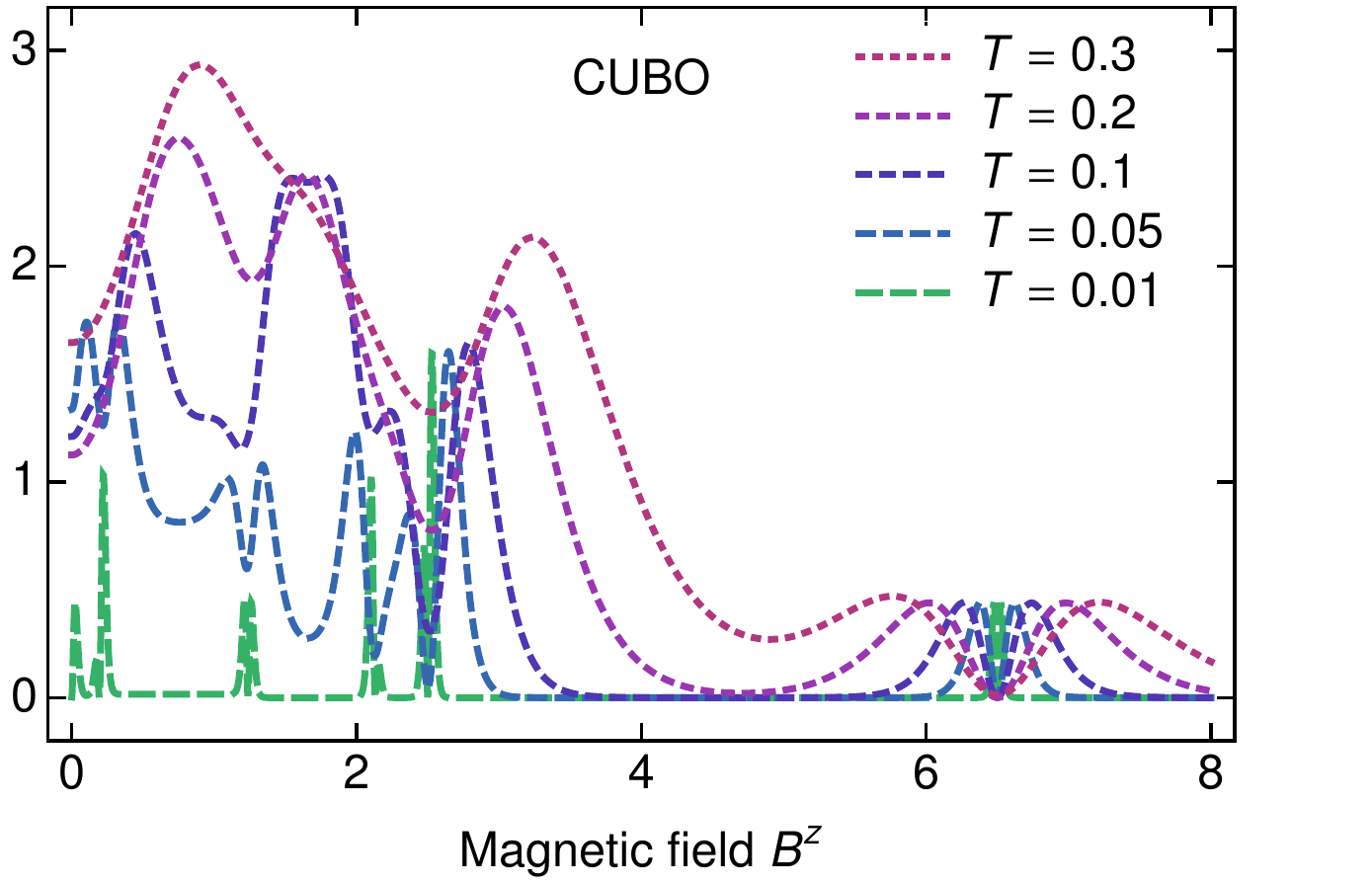}}
\caption{
\label{fig:ff6} (Color online)
Variation of thermodynamic entities as a function of 
external magnetic field for the AFM case of 13-atom ICO (left panel) and 
CUBO (right panel) with spin-1/2. The top
and bottom panels show the variation of entropy $S$ and specific heat $C$
with respect to the magnetic field $B^z$, respectively, at several temperatures $T$.
}
\vspace{0.5cm}
\ibox{\includegraphics[scale=0.57]{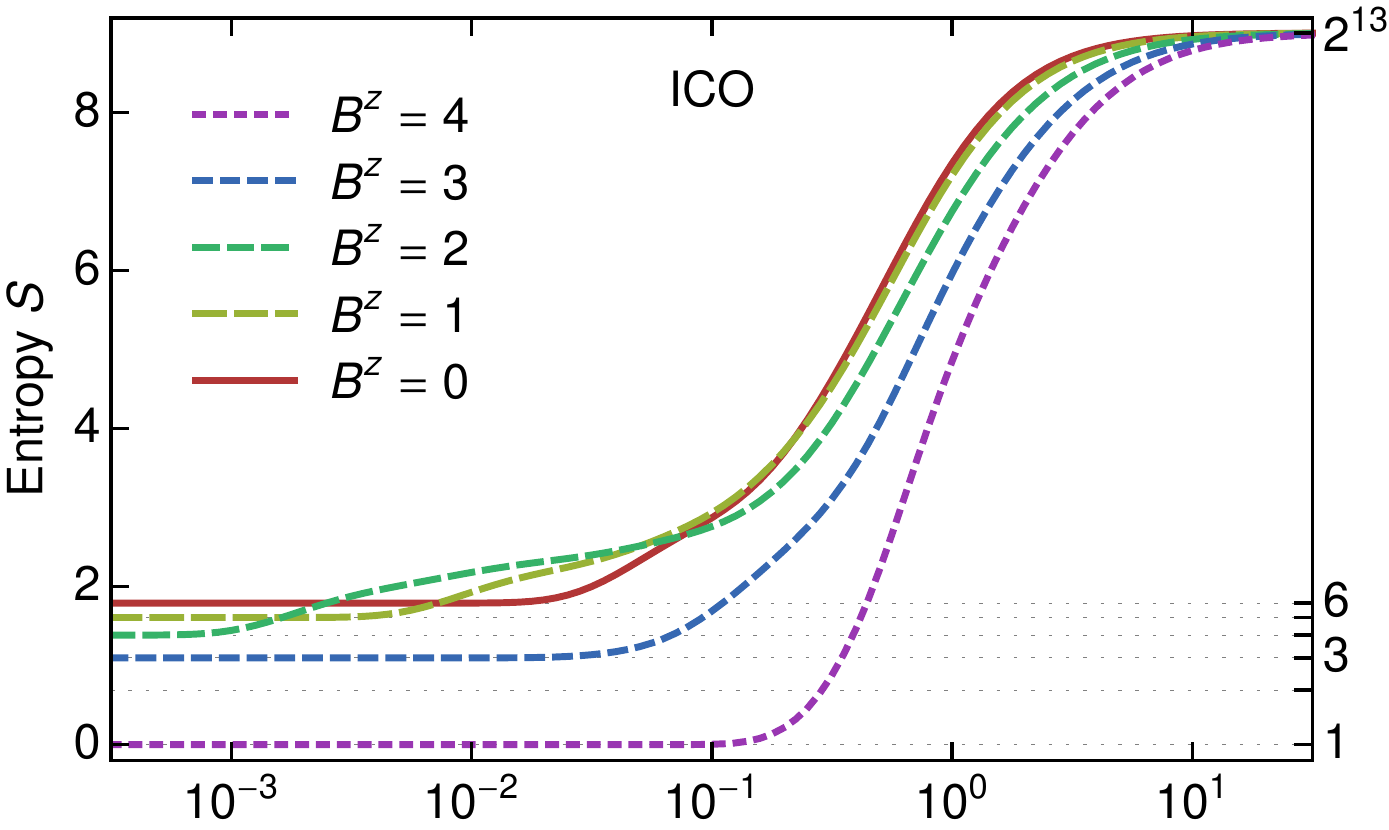}}
\ibox{\includegraphics[scale=0.57]{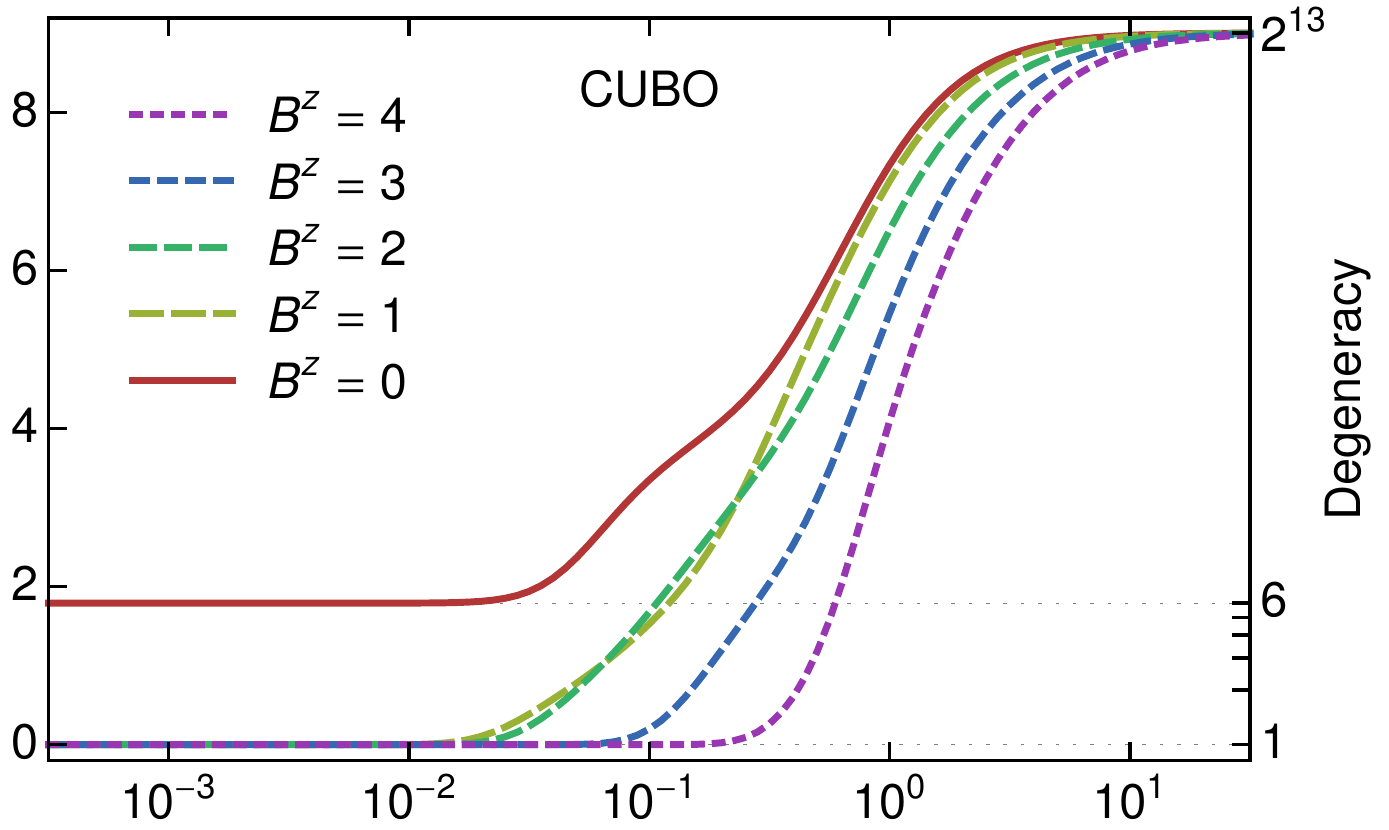}}
\ibox{\includegraphics[scale=0.57]{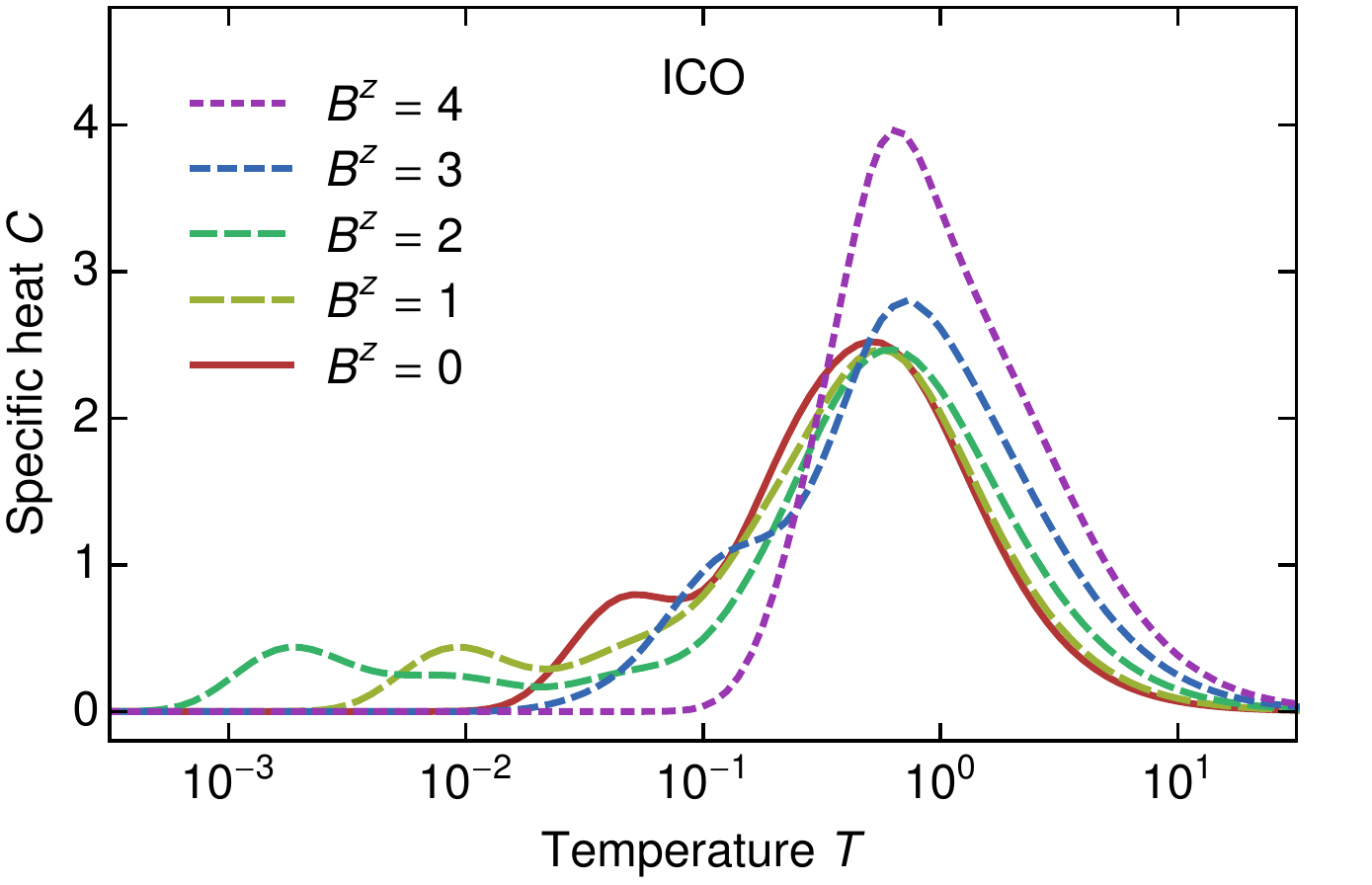}}
\ibox{\includegraphics[scale=0.57]{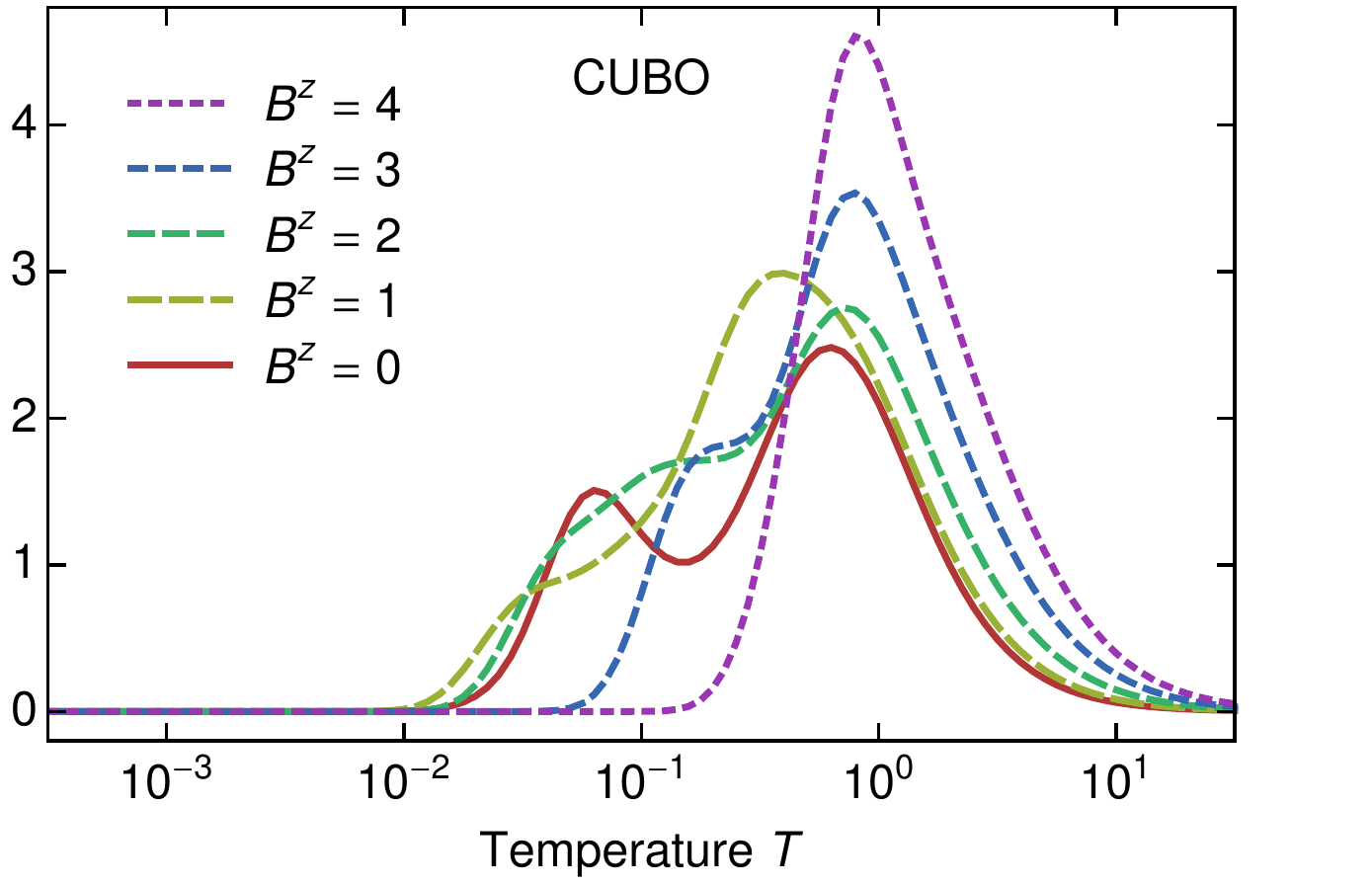}}
\caption{
\label{fig:ff7}(Color online) 
Entropy $S$ (top) and specific heat $C$ (bottom) as a function of temperature (in units of $|J|$)  
for the AFM case of 13-atom ICO (left panels) and CUBO (right panels) with 
spin-1/2 for different magnetic fields $B^z$.
}
\end{figure*}
Now we shall study the effect of dipolar interaction on the magnetization of the 13 atom 
ICO with $s=1/2$ in presence of magnetic field. Since dipolar interaction breaks the isotropy of the 
system, dipole-dipole interaction may be an important source of the observed magnetic anisotropy of various magnetic 
materials~\cite{breed-1967}. More over, as dipole-dipole coupling depends only on 
known physical constants and inverse cube of the interatomic distances, understanding the role of dipolar 
interaction on different properties of the molecule will be useful in the studies of molecular structures. For 
our studies we shall consider the Hamiltonian given in Eq.~(\ref{eqn:eq3}). Using the exact diagonalization 
technique we calculate the magnetic properties of the above mentioned system. 
Figure~\ref{fig:ff5} shows the variation of magnetization as a function of the 
magnetic field for different reduced dipolar interaction strengths,
\begin{equation}
\omega = \frac{\mu_0}{4\pi } \frac{(g\mu_{\mathrm B})^2}{|\vec r_{0i}|^3 |J|},
\end{equation}
$\omega = 0$, 0.025, 0.05 and 0.1 for the AFM case, where $|\vec r_{0i}|$ denotes the shell radius.
For $\omega=0$ we find a reversed central spin $\vec s_0$ with negative hysteresis, as long as 
$B^z/|J'| < 13/2$. 
For larger fields the central spin flips into field direction.
Table~\ref{tab:table-dipole} lists the values of center $s_{0}^z$ 
and surface spin $s_{i}^z$ magnetizations for $\omega = 0$ at different values 
of $S^z$, given by
\begin{equation} \label{eq:s0si}
\langle s_0^z \rangle = -\frac{S^z}{2(S^z+1)}, \hspace{1cm} \langle s_i^z \rangle = \frac{S^z-\langle s_0^z \rangle}{12}.
\end{equation}

At finite values of $\omega$, the magnetization of the surface 
atoms (the light gray curve in Fig.~\ref{fig:ff5})
behave differently depending on their position which indicates that 
the dipolar interaction has a strong impact on the magnetization of these spins. 
The spins of the top and bottom atoms of the cluster (see left panel of 
Fig.~\ref{fig:ico}) are strongly modified (orange curves) compared to the other surface 
spins (light gray) even at very small values of $\omega$, while
the magnetization of the center atom is nearly unaffected by the change in $\omega$ values.   
On the other hand, for the FM case, $\omega$ has no influence on the magnetization of center 
or surface spins. 
\begin{figure*}
{}\hfill
\ibox{\includegraphics[scale=0.50]{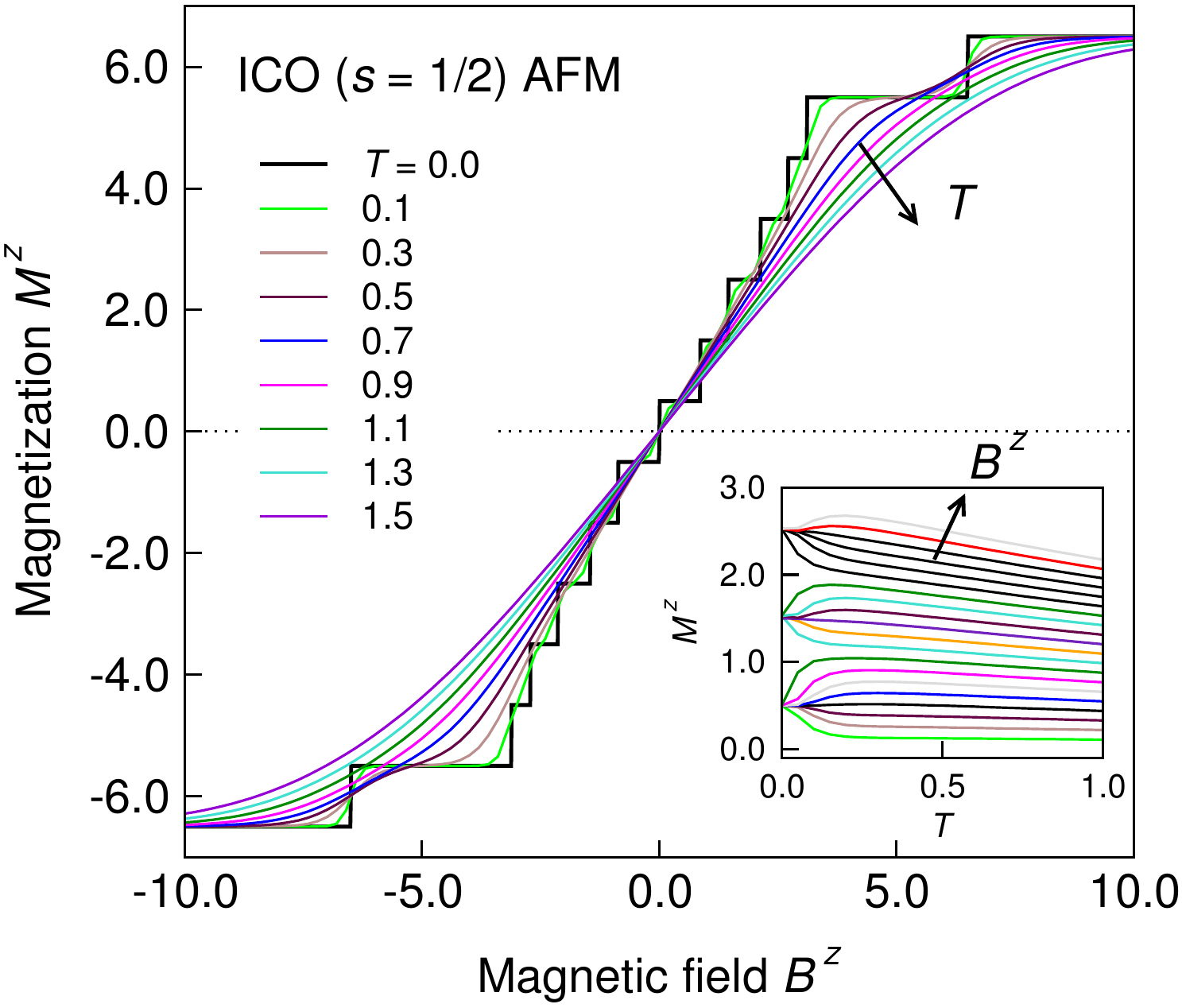}}
\hfill\hfill
\ibox{\includegraphics[scale=0.50]{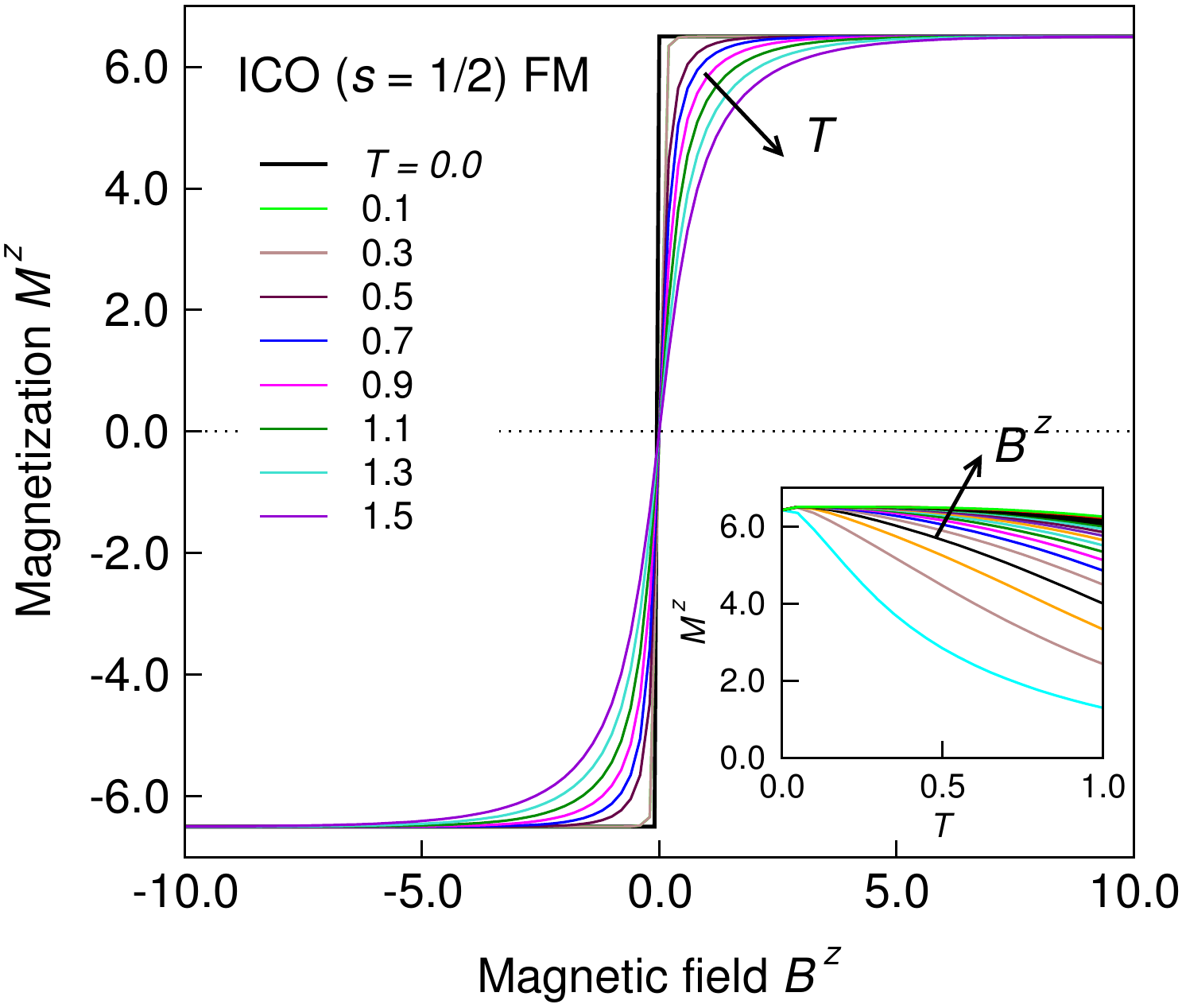}}
\hfill{}
\caption{
\label{fig:ff8}(Color online)
Magnetization as a function of magnetic field measured 
in units of $|J|$ at various temperatures for the AFM (left) and FM (right) cases 
of 13-atom ICO. With increase in temperature (arrows), the plateaus 
start to vanish for the AFM case of 13-atom ICO. The insets for both AFM and FM cases 
shows the variation of magnetization with respect to temperature for several values of 
magnetic fields (arrows). 
}
\end{figure*}

\begin{figure}[b]
{}\hfill
\ibox{\includegraphics[scale=0.17,clip,bb=100 0 820 700]{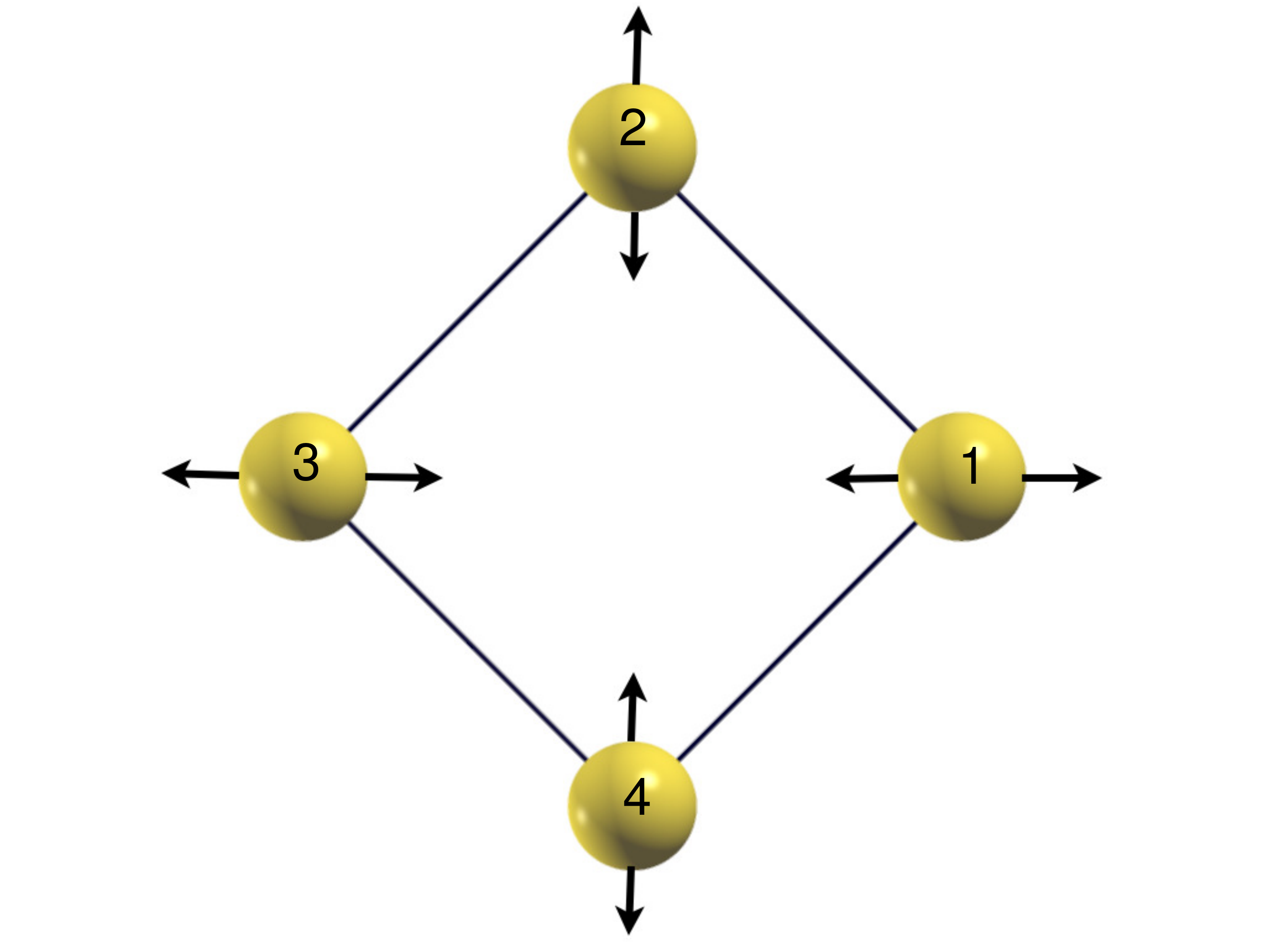}}
\hfill\hfill
\ibox{\includegraphics[scale=0.17,clip,bb=100 0 820 700]{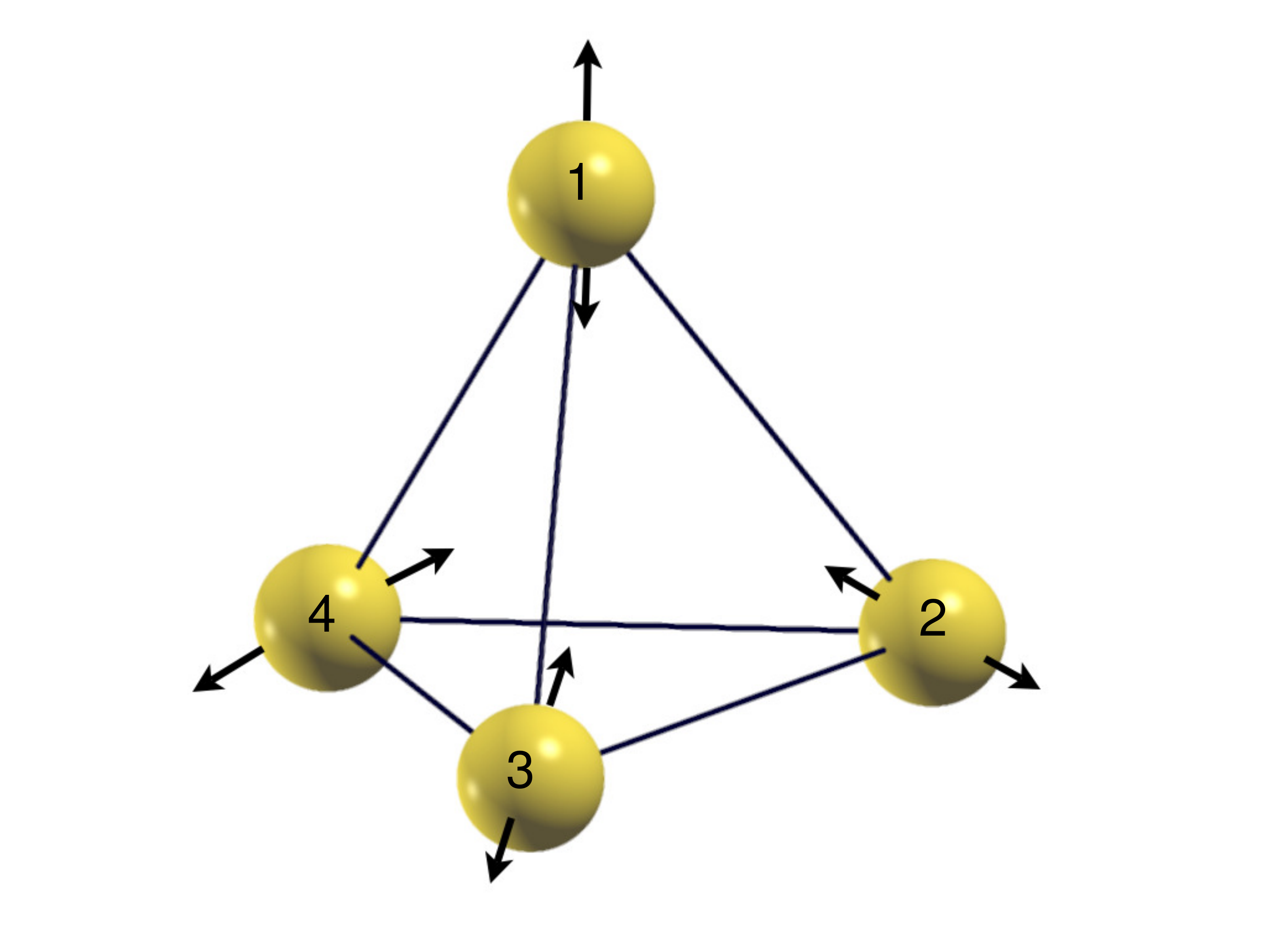}}
\hfill{}
\caption{
\label{fig:sqtetra}(Color online)
Schematic picture of a planar square (left panel) and 
a tetrahedron (right panel). The double arrows indicate the radial anisotropy axes for $s$ = 1. 
For the square case, the magnetic field is perpendicular to the plane 
and for the tetrahedron, it is aligned parallel to $\vec{r}_1$.}
\end{figure}
In order to understand the impact of frustration on the spin configuration 
of the clusters with different geometries, the spin-spin correlation function 
for ICO and CUBO are also calculated. These are directly connected to the 
magnetic structure factor by a Fourier transformation, which in principle 
can be measured experimentally by e.g. neutron scattering techniques. 
Nevertheless, we are not aware of such experiments on clusters. 
The correlation function at finite temperature can be defined as 
\begin{equation}\label{eqn:eq6}
\langle \vec{s}_i\cdot\vec{s}_j \rangle = \frac{\mathrm{Tr}\,e^{-\beta \mathcal H} \, \vec{s}_i\cdot\vec{s}_j}
{\mathrm{Tr} \, e^{-\beta \mathcal H}}
\end{equation}
where, $\beta=1/T$ with the temperature measured in the units of Boltzmann constant. However, at $T=0$ 
the correlation functions  are calculated 
from the eigenvectors obtained from the exact diagonalization of the Hamiltonian 
in Eq.~(\ref{eqn:eq7}) directly. 
The distance dependence of the correlation function at zero temperature  are 
plotted in Fig.~\ref{fig:ffcorr}. 
The correlations for the FM case are found to be the same 
for both geometries, whereas for AFM interactions, we obtain different correlation functions
 for ICO and CUBO, suggesting the existence of frustration in the system. 
However, the ICO seems to be less 
frustrated with respect to the CUBO, as a regular $+-+-$ oscillation is found for 
the ICO, while the CUBO exhibits an irregular and smaller correlations in the 3rd and 4th 
neighbor shell. The ground state correlation functions for the classical 
limit s$\rightarrow$$\infty$ are also calculated for these clusters. 
While comparing the classical and quantum spin correlation functions 
it reveals that the ground state correlation functions for the ICO with $s$=1/2 shows a similar 
qualitative trend compared to the classical case while for the CUBO the correlation 
functions show large deviations in the third and fourth neighbors relative to the 
corresponding classical case. This trend indicates that the CUBO has stronger effect of 
frustration compared to the ICO.\\

In addition, thermodynamic quantities such as entropy $S$ and 
specific heat $C$ are calculated in the absence of dipolar or uniaxial terms in Hamiltonian 
for AFM interactions in the ICO and CUBO  as a function of magnetic field at 
several temperatures, which are shown in Fig.~\ref{fig:ff6}. Sharp peaks at low 
temperature are observed for the AFM case as the magnetic field is changed. 
This is due to the fact that the thermal fluctuation is enhanced at those 
magnetic fields where level crossing occurs. 
With increasing temperature, however, a larger number of states from each $S^z$ sector contributes 
to the thermodynamics, thereby smearing the peaks of the entropy $S$ is observed. Similar explanation can 
be given for the behavior of the specific heat with respect to the magnetic field 
at various temperatures. 
For the FM case (not shown), however, only the maximum $S^z$ block matrix has the lowest energy 
for all magnetic fields. In other words, all eigenvalues are simply scaled with magnetic field 
and thus trivial features observed in the same thermodynamic quantities like $S$ and $C$  
and therefore they are not plotted in the present paper. The thermodynamic observables for the 
antiferromagnetic interactions as a function of temperature for both cluster geometries are shown in 
Fig.~\ref{fig:ff7}. 
The peak in the specific heat curve as a function of temperature at $T \approx 1$ 
marks the classical excitations in the system. However, both systems also have pure quantum excitations
from the low lying energy levels at lower temperatures, visible as additional peaks in the specific heat 
and plateaus in the entropy. For example, in the case of ICO at $B^z = 2$, these excitations are 
at a very low temperature $T\approx 10^{-3}$ and stem from the very small energy gap $\Delta E_1 = 0.0045$
in the $S^z = 5/2$ sector, see Tab.~\ref{tab:table2}. A similar behavior is observed at $B^z = 1$, where
the maximum at $T\approx 10^{-2}$ comes from the small energy gap $\Delta E_1 = 0.022$ in 
the $S^z = 3/2$ sector.

Figure~\ref{fig:ff8} shows the variation of magnetization as a function of the magnetic field at different
temperatures for AFM (left) and FM (right) interactions, respectively. It shows 
that for both interactions, the magnetization is smeared out with increasing 
temperature. The insets in Fig.~\ref{fig:ff8} (left and right) show the variation of 
magnetization with respect to the temperature at different magnetic fields, which shows that quantum
effects vanish at around $T\approx 0.1$ in the AFM case and that
the total magnetization tends to decrease with increasing temperature. 
For CUBO, a similar variation of magnetization as a function of magnetic field 
is observed at several temperatures.        

\begin{figure*}
\begin{center}
\includegraphics[scale=0.5]{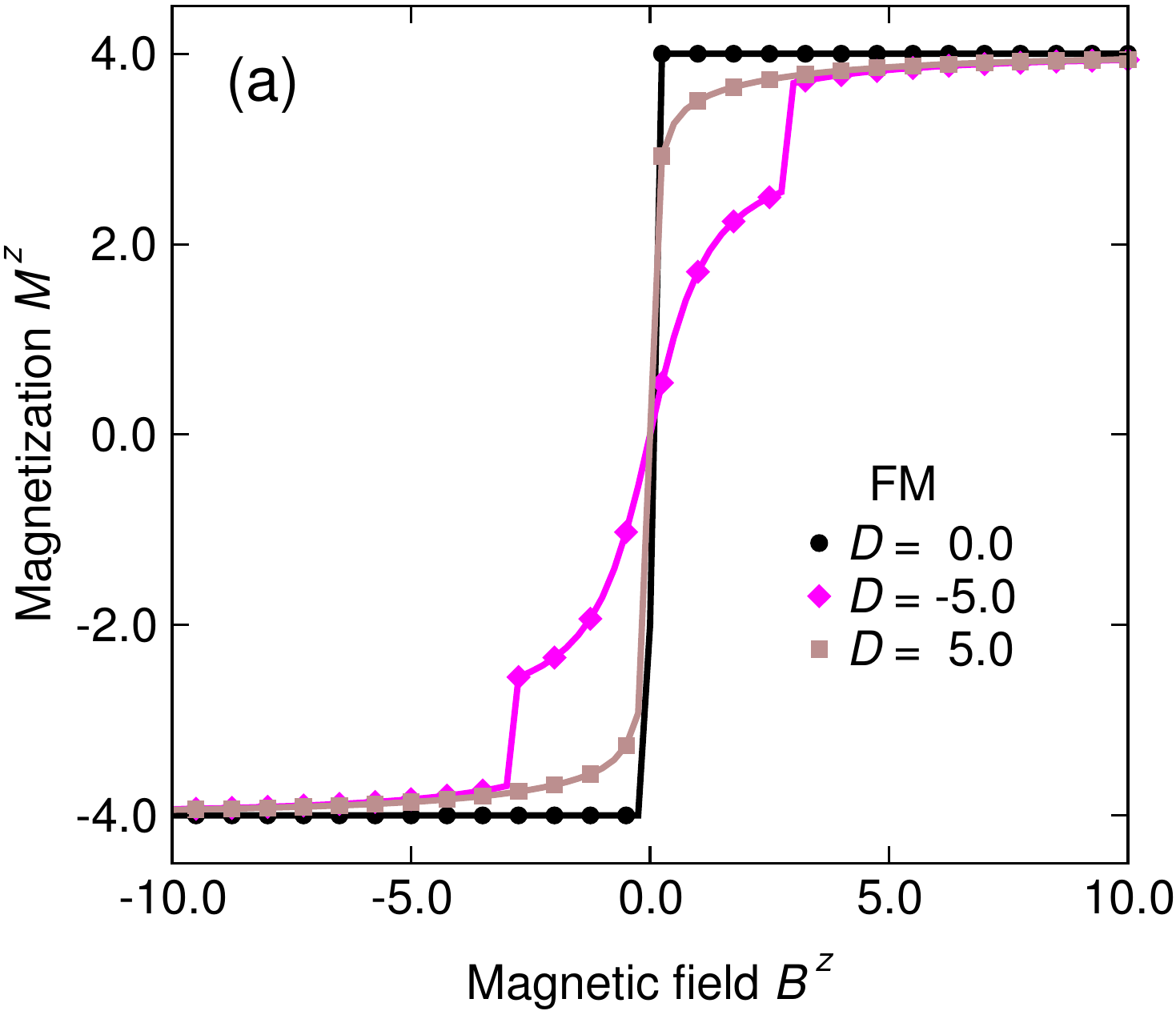}%
\hspace{1cm}%
\includegraphics[scale=0.5]{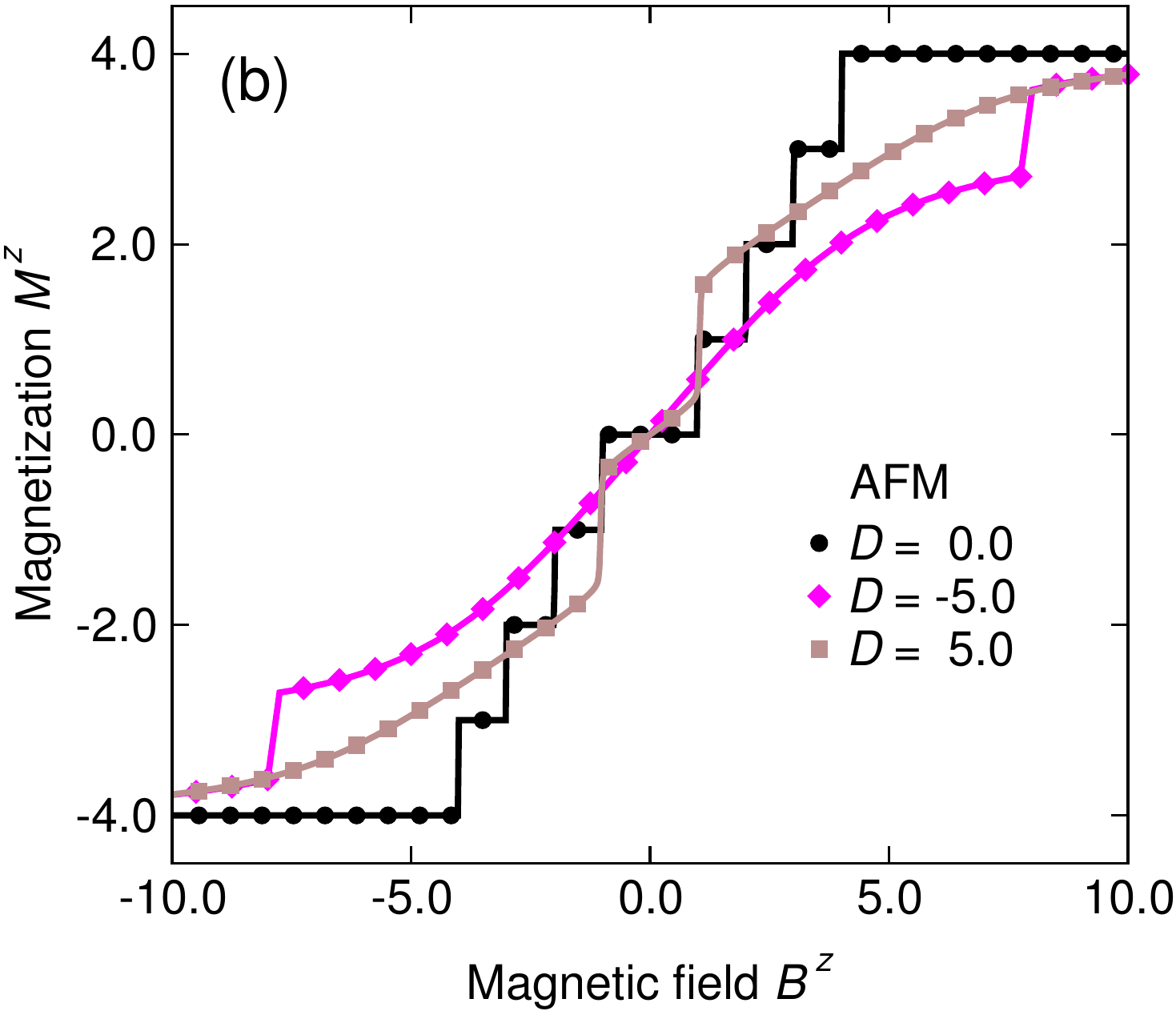}
\vspace{0.7cm}
\\
\includegraphics[scale=0.5]{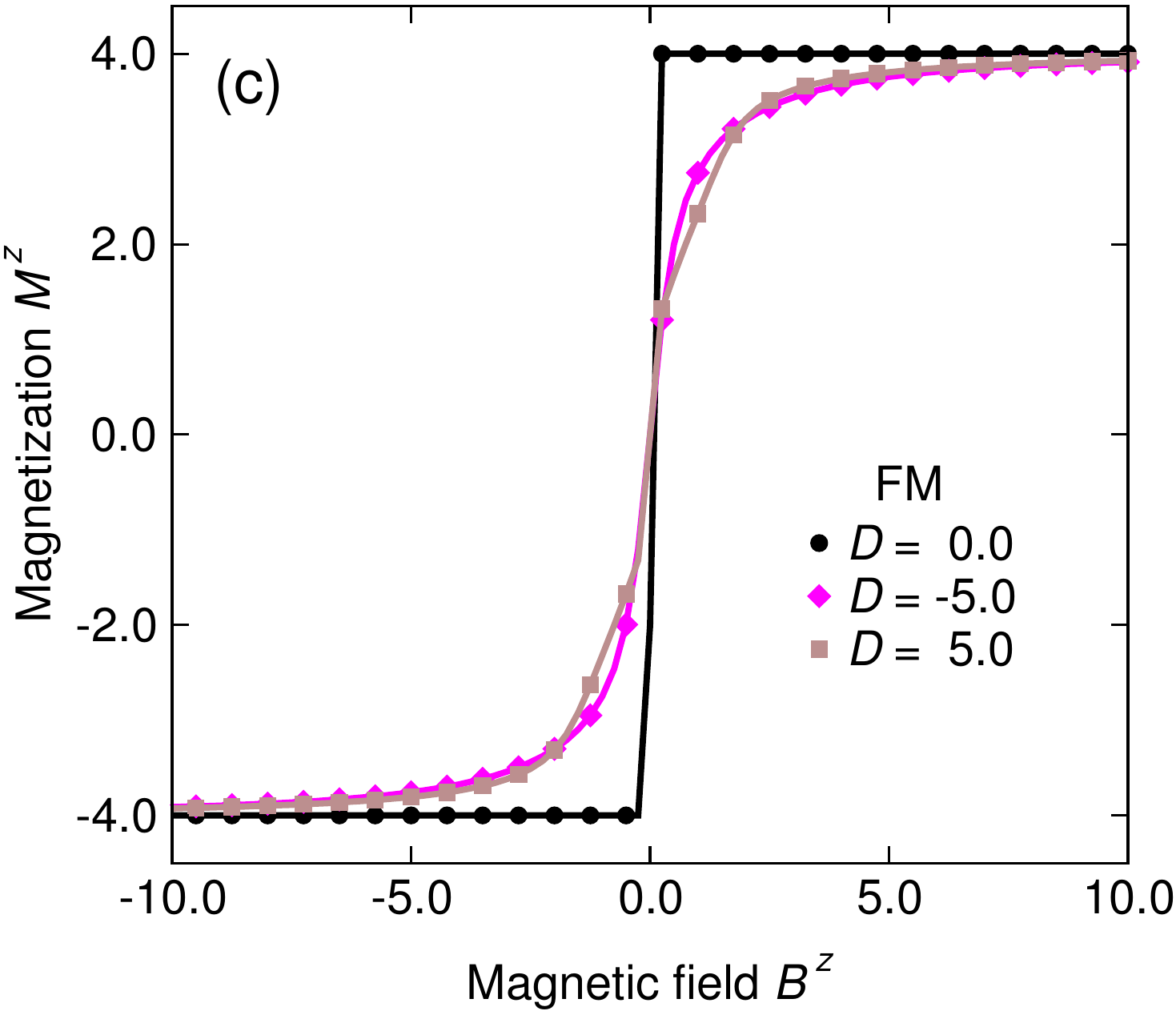}%
\hspace{1cm}%
\includegraphics[scale=0.5]{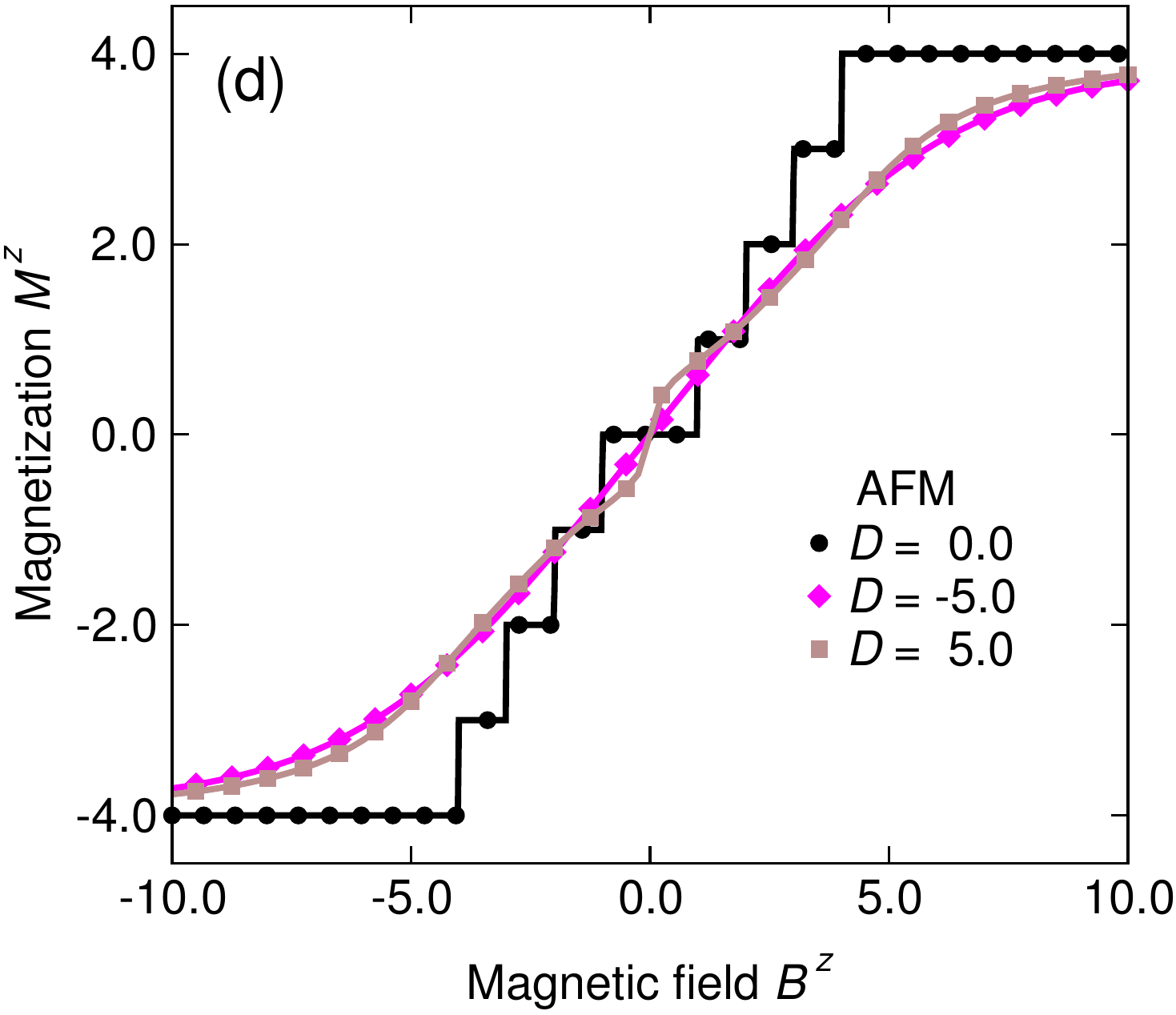}
\end{center}
\caption{
\label{fig:ff11} (Color online) The effect of anisotropy on magnetization as a function of 
external magnetic field (measured in units of $|J|$) for the spin-1 tetrahedron (a,b) and 
square (c,d). The results for FM and AFM interactions are shown in the left and right panels, respectively. 
The black line (circles) shows the variation of magnetization as a function of the 
magnetic field. The curves with squares (brown) and diamonds (magenta) show the variation of 
the same quantity for positive and negative $D$.
}
\end{figure*}

\begin{figure*}
\begin{center}
\includegraphics[scale=0.5]{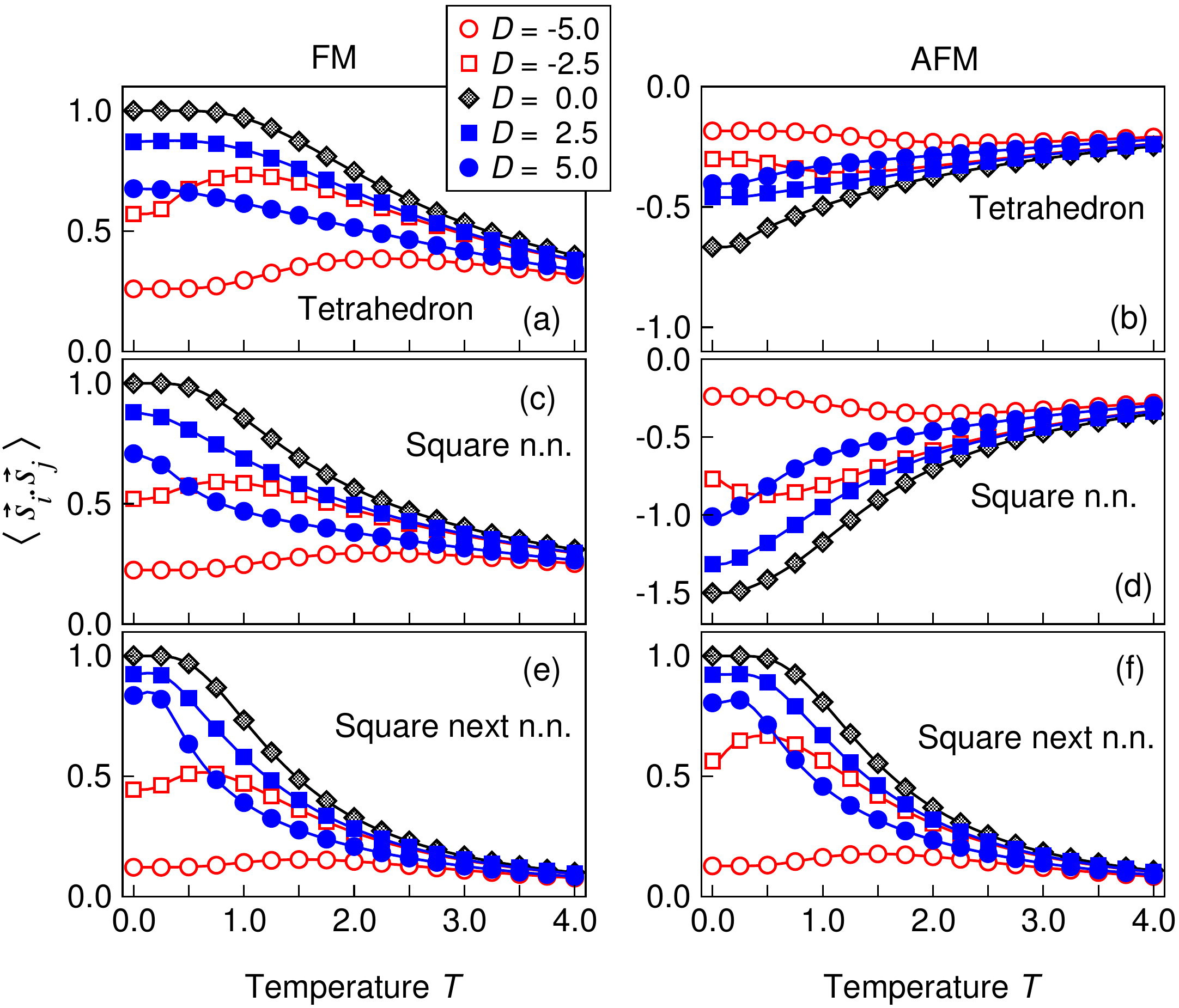}
\end{center}
\caption{
\label{fig:ff13} (Color online) Variation of the spin-spin correlation function with 
temperature (expressed in units of $|J|$) for the FM and AFM spin-1 tetrahedron (a), (b) and 
square (c)-(f). Diamonds, open symbols and filled symbols represent the correlation 
functions $\langle \vec{s}_i\cdot\vec{s}_j \rangle$ for zero, negative and positive 
anisotropy constants $D$, respectively.}
\end{figure*}

\section{4-atom cluster with spin-1}
Now we present results for 4-atom clusters with $s=1$ and uniaxial 
anisotropies (see Eq.~(\ref{eqn:eq4})).
It may be noted that such anisotropies only give a constant for $s$ = 1/2. 
For a spin-1 system, the total Hamiltonian 
in the presence of local uniaxial anisotropy axes $\vec e_i$ reads
\begin{equation}\label{eqn:eq5}
\mathcal H_{4} = - \sum_{i < j} J_{ij} \, \vec{s}_{i}\cdot\vec{s}_{j} - B^z S^z
 - \sum_{i} D_i \, (\vec{e_i} \cdot \vec{s_{i}})^{2},
\end{equation}   
where $D_i$ are the local uniaxial anisotropy constants and $\vec{e_i}$ are the easy axes 
compatible with the symmetry of the system \cite{sahoo-2010}. 
We have previously studied the structural and magnetic properties of small
transition metal clusters with more emphasis on the magnetic anisotropy using the density 
functional theory (DFT)~\cite{sahoo-2010}, where the energies obtained from DFT calculations were 
fitted by using a classical Heisenberg Hamiltonian. The investigations presented here can be 
viewed as a continuation of the previous work in the sense that we perform exact 
diagonalization of a corresponding quantum spin Hamiltonian to study the system.

There has been several studies related to the magnetic and thermodynamic properties 
for spin-1 clusters~\cite{kamieniarz-1992,li-2008} through Heisenberg model. However, studies including 
the effect of local uniaxial anisotropy on several properties of clusters is still 
limited~\cite{klemm-2008}. In the present work, 
we have studied the influence of radial anisotropy on the magnetic properties and 
temperature-dependent correlation functions for the spin-1 tetrahedron and square 
as shown in Fig.~\ref{fig:sqtetra}. A regular tetrahedron (symmetry group $T_d$) 
consists of four triangular faces, whereby the
triangles meet at each vertex and are equilateral. A square is a regular quadrilateral
with $D_4$ symmetry. 

In the presence of radial anisotropy, the Hamiltonian is modified to a form as 
represented in Eq.~(\ref{eqn:eq5}), where $D_i=D$ is the anisotropy constant and $\vec{e}_i$ 
are the easy axes which  differs for each spin. In Fig.~\ref{fig:sqtetra} the 
anisotropy axes (double arrows) pointing 
into the radial directions are shown for the square and tetrahedron. 
For the 4-atom spin-1 cluster, 
the Hamiltonian matrix is of dimension $3^4 \times 3^4$. For $D=0$, the Hamiltonian matrix 
can be decomposed into 9 block matrices with $S^z = -4, -3,\ldots, 4$.
However, at finite $D$ the block matrix structure is destroyed as the uniaxial anisotropy term 
does not commute with total $S^z$  
and the whole Hamiltonian matrix has to be diagonalized.

The presence of anisotropy $D$ results in a different qualitative behavior of magnetization 
as a function of the external magnetic field as shown in Fig.~\ref{fig:ff11} for the 
tetrahedron (a-b) and the square (c-d). 
In the absence of magnetic anisotropy ($D = 0$), we obtain a single step in the magnetization 
for FM exchange interaction and $9$ plateaus for AFM exchange interaction 
for both clusters, the tetrahedron and the square. 
The presence of anisotropy leads to the smearing of magnetization with respect to magnetic field 
for FM and AFM interactions (see the squares and diamonds) of square and tetrahedron, since 
$\mathcal H_\mathrm{ani}$ mixes the eigenstates of $\mathcal H_\mathrm{0}$ with different total spin values. 
In particular, for tetrahedron geometry, we observe differences in the magnetization 
for positive and negative values of $D$, whereas for the square geometry, 
the change of sign in the anisotropy does not affect the magnetization significantly. This shows the 
dependence of anisotropy on the structural symmetry, which has been observed earlier 
for 13-atom clusters through Monte Carlo simulations~\cite{hernandez-2005}.    

In addition, we have calculated the temperature-dependent correlation functions, as defined 
in Eq.~(\ref{eqn:eq6}),  for the 
4-atom clusters with spin-1 in the presence of anisotropy.
The variation of the nearest-neighbor correlation functions with respect to temperature 
has been plotted in Fig.~\ref{fig:ff13} for the spin-1 tetrahedron (a-b) and square (c-f) with 
different anisotropy constants for the FM (left panel) and AFM (right panel) case. For both interactions, 
it has been observed that the anisotropy modifies the correlation function significantly at 
low temperatures. The 
correlations are positive in the FM cases and negative in the AFM case. 
For $D\geq0$ the correlations decrease with temperature as expected.
However, for negative $D$ the correlations are reduced at low temperatures due to the quantum effects, 
as can be seen most easily for the square: Here the classical ground state would be 
a FM/AFM state in the direction perpendicular to the plane, as in this case all couplings 
are satisfied and the spin directions are all perpendicular to the anisotropy axes as required. 
However, $\mathcal H_\mathrm{ani}$ introduces spin-flip process in the system and as a consequence 
ground state becomes the linear combination of states with different magnetization. 
Hence a strong reduction in correlations is observed in presence of anisotropy term at low temperatures.  
This reduction of correlations becomes negligible at higher temperatures, and correlations decrease as usual.
While the nearest neighbor correlations in the AFM square (Fig.~\ref{fig:ff13}d) are negative, 
the second nearest neighbor correlations are positive, as the square is not frustrated.
\begin{table}[t]
\caption{
\label{tab:ICO1}Size of block matrix, lowest energy eigenvalues $E_0$, $\Delta E_1$, and degeneracies $K_0$, $K_1$, for 
different $S^{z}$ for 13-atom AFM ICO with spin-1.
}
{
\begin{tabular*}{0.9 \columnwidth}{@{\extracolsep{\fill}} cccccc}
\hline
\hline
\vspace{-0.3cm}\\
 $|S^{z}|$ & Matrix size & $E_0/|J|$ & $K_0$ & $\Delta E_1/|J|$ & $K_1$\\
\vspace{-0.3cm}\\
\hline
13 & 1 & $+42$ & 1 & $-$ & $-$ \\
12 & 13 & $+29$ & 1 & $4.763932$ & 3 \\
11 & 91 & $+17$ & 1 & $4.763932$ & 3 \\
10 & 442 & $+10.763932$ & 3 & $1.236068$ & 5 \\
 9 & 1651 & $+5.034063$ & 5 & $0.147456$ & 1 \\
  8 & 5005 & $-0.131753$ & 4 & $0.061759$ & 3 \\
  7 & 12727 & $-4.663902$ & 4 & $0.032182$ & 5 \\
  6 & 27742 & $-8.608201$ & 3 & $0.035204$ & 4 \\
  5 & 52624 & $-11.932667$ & 4 & $0.040381$ & 4 \\
  4 & 87802 & $-14.679508$ & 5 & $0.011089$ & 3 \\
  3 & 129844 & $-16.920343$ & 5 & $0.007569$ & 4 \\
  2 & 171106 & $-18.566489$ & 3 & $0.013245$ & 4 \\
  1 & 201643 & $-19.506298$ & 5 & $0.007972$ & 5 \\
  0 & 212941 & $-19.839976$ & 3 & $0.034983$ & 3 \\
\hline
\hline
\end{tabular*}
}
\end{table}

\section{spin-1 Icosahedron}
\begin{figure}
\begin{center}
\includegraphics[scale=0.6]{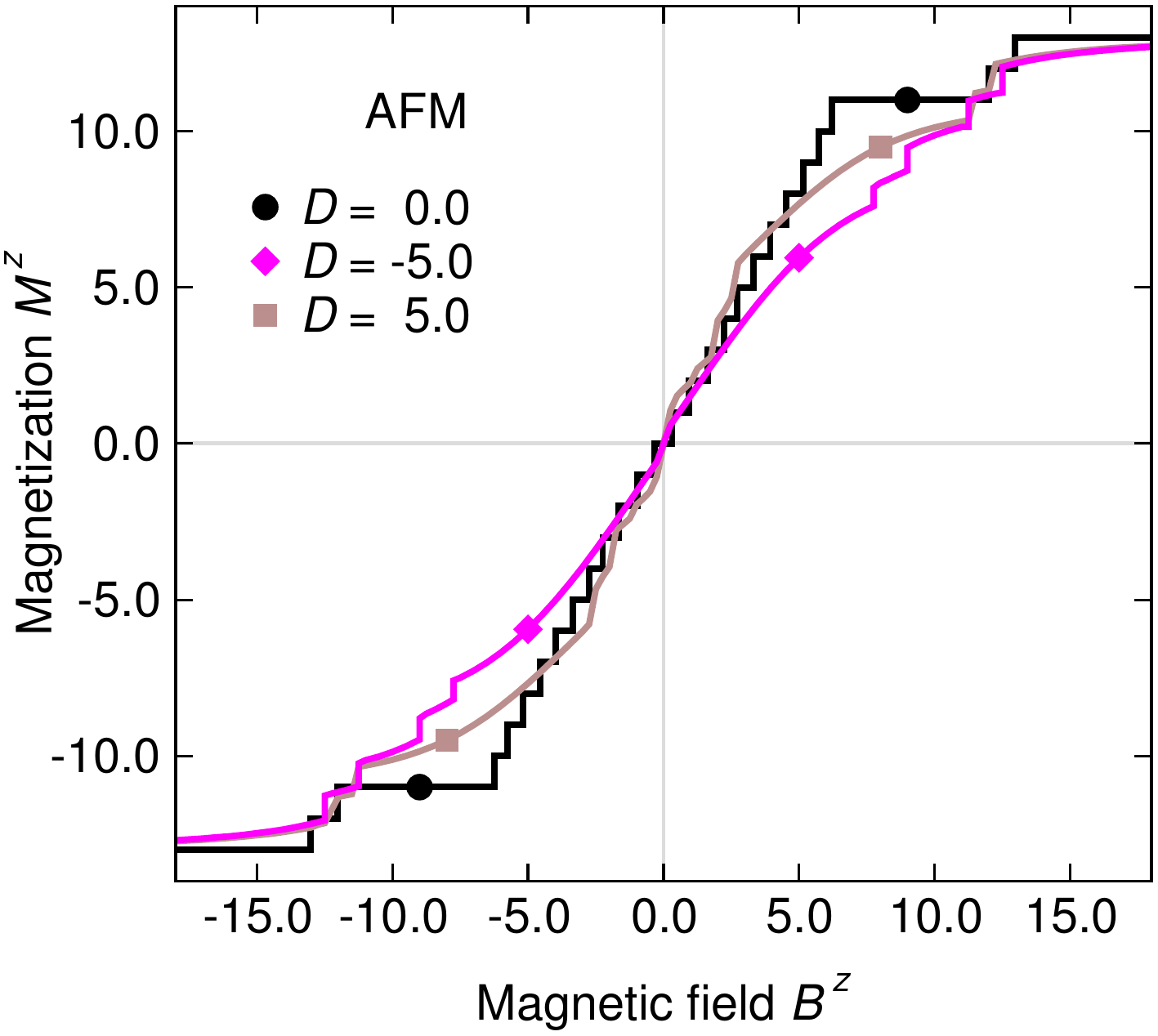}
\end{center}
\caption{
\label{fig:ICO1-hys} (Color online) Variation of magnetization $M^z$ as a function of the
external magnetic field $B^z$ (measured in units of $|J|$) for different values
of the radial uniaxial anisotropy $D$ for the spin-1 AFM icosahedron.
}
\end{figure}

Finally we present the results for the spin-1 icosahedron with 
Hamiltonian described in Eq.~(\ref{eqn:eq7}). Now the  Hamiltonian matrix 
has $3^{13} = 1594323$ columns and rows. For vanishing anisotropy, it can be decomposed into 
block matrices whose sizes are the trinomial numbers $ ({13 \atop k})_2$ for $-13 \leq k \leq 13$, given 
in Tab.~\ref{tab:ICO1} together with the 
lowest eigenvalues of each $S^z$ block. From these results we compute the hysteresis curve shown in 
Fig.~\ref{fig:ICO1-hys} as black line. The curve is similar to the spin-1/2 case, since 
for fields below $|B^z| \lesssim 6$ 
the magnetization varies with nearly constant step size from -11 to 11. The plateaus at $S^z = \pm 11$ mark the 
saturated outer shell, where only the central spin still points antiparallel to the external field. At 
larger fields also $\vec s_0$ aligns with the field in two steps, one from $\langle s_0^z \rangle = -1 {\rightarrow} 0$ 
at $B^z = 12$ and one from $\langle s_0^z \rangle = 0 {\rightarrow} 1$ at $B^z = 13$. This behavior already shows 
characteristics of the classical limit, in which the steps vanish and the magnetization varies continuously with field. 
Nevertheless, also in this limit we find a plateau at $M^z=11$ with reversed central spin, which rotates towards 
the field direction in the range $11<B^z<13$.

The ground state spin correlation functions for the AFM case of ICO with spin-1 
are plotted in Fig.~\ref{fig:ffcorr} (see the filled squares). This shows again 
similar qualitative behavior to the spin-1/2 case 
and show antiferromagnetic order in the AFM case, interestingly now the anti-correlation 
between the third neighbor atoms is larger than the nearest neighbor value. 
This shows that the correlation approaches towards the classical limit $-1$, as shown in Fig.~\ref{fig:ffcorr} 
(the dashed line). On the other hand, for the FM case, the correlation functions for all shell 
neighbors become one for the ICO.

Additionally, we have calculated the hysteresis for the icosahedron with 
radial uniaxial anisotropy according to Eq.~(\ref{eqn:eq5}). For these calculations 
it was again necessary to work with the whole Hamiltonian matrix, as the 
anisotropy term does not commute with the interaction term and thus 
destroys the block structure of $\mathcal{H}$.
Nevertheless, we could calculate the smallest eigenvalues and eigenvectors using 
a Lanczos scheme. The resulting curves are shown in Fig.~\ref{fig:ICO1-hys} as 
magenta ($D=-5$) and brown ($D=+5$) lines, they are similar to the results for 
the tetrahedron with anisotropy shown in Fig.~\ref{fig:ff11} a,b. 


\section{Summary and outlook}
We have employed the exact diagonalization technique for small spin clusters 
with 4 and 13 vertices using the quantum-mechanical nearest-neighbor Heisenberg model and calculated 
the full energy spectrum numerically as well as the ground state energies of the 13-atom systems analytically. 
The magnetic and thermodynamic properties as well as the spin-spin correlation functions are derived from
these results. The ground state magnetization shows discontinuities accompanied by a magnetization plateau 
as a function of magnetic field for the antiferromagnetic exchange interaction. These magnetization 
plateaus vanishes for temperatures around $T \gtrsim 0.1 |J|$.  
The ground state correlations suggest that the icosahedron is less frustrated than the cuboctahedron, 
since a regular $+-+-$ oscillation is found for the ICO, while the CUBO has
irregular correlations with smaller values in the 3rd and 4th neighbor shell, see Fig.~\ref{fig:ffcorr}.

We have shown that the dipolar interaction plays a significant role for the magnetization 
in case of AFM interaction of the ICO. Our investigations show that dipolar interactions 
have a strong influence on the magnetization of surface atoms in an 
external magnetic field, while the field-dependent magnetization of the center 
atom remains nearly unchanged by the dipolar interactions. 

The field dependence of magnetization and temperature dependence of correlation 
function on tetrahedron and square for $s=1$ indicates that the influence of radial anisotropies on the 
magnetic properties strongly depends on the structural symmetry of the cluster. 

Finally, we have investigated the 13-atom icosahedron with $s=1$,
which involves quite large Hamiltonian matrices of dimensions $3^{13} \times 3^{13}$ that cannot be
decomposed into smaller block matrices if the local uniaxial anisotropy axes are present. 
Using Lanczos methods we calculated the lowest eigenvalues and corresponding 
eigenvectors of these large matrices, determined correlation functions and 
hysteresis curves and compared these results to the spin-1/2 cases.

Regarding any comparison of this exact diagonalization calculation with 
experiments one has to first of all note that this requires tiny negative exchange coupling 
to observe the quantum effects (such as steps) in reasonable magnetic field. 
Nevertheless, there exists few examples in nature such as Mn$_{12}$-acetate and Mn$_{4}$-dimer 
molecules embedded in organic ligands, which fulfills this condition, see for instance 
Refs.~\cite{gatteschi-2006, dorfer-2002}. Another interesting point would be a collection of quantum clusters, 
showing long-range order at low temperatures. For such calculations, exact diagonalization
results can be used as a basis. This is left for future studies. 


\begin{acknowledgments}
One of the authors (A. H.) thank Bj\"orn Sothmann for helpful discussions.
\end{acknowledgments}

\end{document}